\documentclass[preprint,12pt]{elsarticle}

\usepackage[hidelinks]{hyperref}
\usepackage{graphicx}
\usepackage{subcaption}
\usepackage{amssymb}
\usepackage{amsmath}
\usepackage{multirow}
\usepackage[utf8]{inputenc}
\usepackage{ifthen}
\usepackage{cleveref}
\usepackage{listings}
\usepackage[ruled,vlined]{algorithm2e}
\usepackage{booktabs}

\usepackage{xcolor}
\usepackage{xargs}
\usepackage[colorinlistoftodos,textsize=footnotesize]{todonotes}

\newcommandx{\unsure}[2][1=]{\todo[linecolor=red,backgroundcolor=red!25,bordercolor=red,#1]{#2}}
\newcommandx{\change}[2][1=]{\todo[linecolor=blue,backgroundcolor=blue!25,bordercolor=blue,#1]{#2}}
\newcommandx{\info}[2][1=]{\todo[linecolor=OliveGreen,backgroundcolor=OliveGreen!25,bordercolor=OliveGreen,#1]{#2}}

\newcommand{\ten}[1]{\ensuremath{\mathbf{#1}}}

\newcommand{\code}[1]{\lstinline{#1}}
\newcommand{\codek}[1]{\lstinline[keywordstyle=\color{black}]{#1}}

\DeclareMathOperator*{\argmin}{arg\,min}

\usepackage{lineno}
\usepackage{tikz}
\usetikzlibrary{arrows.meta, patterns}
\usetikzlibrary{shapes, calc,quotes,angles}
\tikzset{%
  >={Latex[width=2mm,length=2mm]},
  base/.style = {
    rectangle, rounded corners, draw=black, minimum width=4cm, minimum
    height=1cm, text centered, font=\sffamily},
  decision/.style = {
    diamond, draw, rounded corners, fill=blue!20, minimum width=3cm, minimum height=0.5cm, text centered, font=\sffamily},
  activityStarts/.style = {base, fill=blue!30},
  startstop/.style = {base, fill=red!30},
  activityRuns/.style = {base, fill=green!30},
  process/.style = {
    base, minimum width=2.5cm, fill=orange!15, font=\ttfamily},
}
\journal{}
\begin{document}

% set showimages to 0 to shutoff all the images
\def\showimages{1}
%%%%%%%%%%%%%%%%%%%%%%%%%%%%%%5

\begin{frontmatter}

  \title{Algorithms for uniform particle initialization in domains
  with complex boundaries}

  \author[IITB]{Pawan Negi\corref{cor1}}
  \ead{pawan.n@aero.iitb.ac.in}
  \author[IITB]{Prabhu Ramachandran}
  \ead{prabhu@aero.iitb.ac.in}
\address[IITB]{Department of Aerospace Engineering, Indian Institute of
  Technology Bombay, Powai, Mumbai 400076}

\cortext[cor1]{Corresponding author}

\begin{abstract}

  Accurate mesh-free simulation of fluid flows involving complex boundaries
  requires that the boundaries be captured accurately in terms of particles.
  In the context of incompressible/weakly-compressible fluid flow, the SPH
  method is more accurate when the particle distribution is uniform. Hence,
  for the time accurate simulation of flow in the presence of complex
  boundaries, one must have both an accurate boundary discretization as well as
  a uniform distribution of particles to initialize the simulation. This
  process of obtaining an initial uniform distribution of particles is called
  ``particle packing''. In this paper, various particle packing algorithms
  present in the literature are implemented and compared. An improved
  SPH-based algorithm is proposed which produces uniform particle
  distributions of both the fluid and solid domains in two and three
  dimensions. Some challenging geometries are constructed to demonstrate the
  accuracy of the new algorithm. The implementation of the algorithm is open
  source and the manuscript is fully reproducible.
\end{abstract}


\end{frontmatter}

% \linenumbers
\section*{Program summary}
\noindent
\textit{Program title}: SPHGeom

\noindent
\textit{Licensing provisions}: BSD 3-Clause

\noindent
\textit{Programming language}: Python

\begin{sloppypar}
  \noindent
  \textit{External routines/libraries}: PySPH
  (\url{https://github.com/pypr/pysph}), matplotlib
  (\url{https://pypi.org/project/matplotlib/}), automan
  (\url{https://pypi.org/project/automan/}),
  ParaView(\url{https://www.paraview.org/download/}).
\end{sloppypar}
\noindent
\textit{Nature of problem}: Particle methods require that complex geometry be
represented accurately when discretized with particles. The particles should
be uniformly distributed inside, outside, and on the surface of the geometry.
A particle packing algorithm is proposed to achieve this. For a fluid flow
past a solid body, the code generates a set of solid particles inside and on
the surface surrounded by fluid particles such that the density is
approximately constant. These particles can be placed anywhere in the main
simulation.

\noindent
\textit{Solution method}: An SPH-based algorithm is proposed where the number
density gradient, a repulsion force, and a damping force are used to move
particles. Particles are constrained near the boundary to move along the
surface. Particles are iteratively projected onto the boundary surface. Once a
desired distribution of particles is achieved, the particles are separated
into interior and exterior particles using the boundary information.  This may
be used directly as an input for particle-based simulation.

\noindent
\textit{Additional comments}: The source code for this repository can be found at
\url{https://gitlab.com/pypr/sph_geom}.

\section{Introduction}
\label{sec:intro}

Smoothed Particle Hydrodynamics (SPH) is a mesh-free numerical method used for
the simulation of continuum mechanics problems. It was first proposed by
\citet{monaghan-gingold-stars-mnras-77} and \citet{lucy77}. Unlike mesh-based
methods, the SPH method discretizes the domain into particles that carry
physical properties. The SPH method has been employed to study a wide variety
of problems. A comprehensive list of schemes can be found in
\citet{liu_smoothed_2010}, \citet{violeau_book_2012}, and
\citet{ye2019smoothed}.

The SPH method approximates functions and derivatives by using a convolution
with a smooth kernel function
\begin{equation}
  f(\ten{x}) = \int_{\Omega} f(\tilde{\ten{x}})
  W(\ten{x} - \tilde{\ten{x}}, h) d \tilde{x} + O(h^2),
  \label{eq:cont-kernel}
\end{equation}
where $\Omega$ is the domain, $W(\ten{x} - \tilde{\ten{x}}, h)$ is a compact
kernel function with properties described in \cite{monaghan-review:2005} and
$h$ is the support radius of the kernel. The domain is discretized into
particles carrying the required properties. This integral is then discretized
as
\begin{equation}
  f(\ten{x}) \approx \sum_{j} f(\ten{x}_j)  W(\ten{x} - \ten{x}_j, h) \frac{m_j}{\rho_j},
  \label{eq:disc-kernel}
\end{equation}
where $\ten{x}_j$ is the position of the $j$\textsuperscript{th} particle and
$V_j = \frac{m_j}{\rho_j}$ is the volume associated with the particle with
mass $m_j$ and density $\rho_j$. This approximation is most accurate when the
underlying particle distribution is
uniform~\cite{sph:basa-etal-2009,kiara_sph_2013}. For example, the particles
could be placed on a uniform Cartesian mesh with constant spacing. However,
the SPH method is Lagrangian and the particles move with the local velocity
field. Many recent developments in SPH involve the use of Particle Shifting
Techniques
(PST)\cite{diff_smoothing_sph:lind:jcp:2009,xu2009accuracy,Adami2013} which
shift the particles towards a uniform distribution. Thus, it is important to
have a uniform distribution of particles.

Consider the case of the two-dimensional flow past a simple non-Cartesian
solid shape like a circular cylinder. The fluid flow occurs outside of the
cylinder. In the SPH method, the fluid is discretized using (fluid) particles.
The circular cylinder is also discretized with (solid) particles. These solid
particles are not merely on the surface of the cylinder but also on its
interior for accuracy of the method. When the simulation starts, we
require a uniform distribution of both fluid and solid particles which
capture the geometry accurately. Clearly, one cannot use a Cartesian mesh of
particles and still describe the cylinder surface accurately.

We must therefore define the term ``uniform'' distribution of particles in
the context of the SPH method. This has been discussed by
\citet{colagrossi2012particle} and \citet{litvinov_towards_2015}. They
explain that a non-uniform distribution would generate a spurious force among
the particles leading to a ``resettlement'' of their spatial distribution.

As discussed in \cite{colagrossi2012particle}, for particles having the same
mass and density, we can estimate the non-uniformity using either,
\begin{equation}
  \Gamma_i = \sum_{j}  W(\ten{x_i} - \ten{x}_j, h) \frac{m_j}{\rho_j},
  \label{eq:gamma}
\end{equation}
where $\ten{x}_i$ refers to the $i$\textsuperscript{th} particle or
\begin{equation}
  \nabla \Gamma_i = \sum_{j}  \nabla W(\ten{x}_i - \ten{x}_j, h) \frac{m_j}{\rho_j}.
  \label{eq:del-gamma}
\end{equation}
where $\nabla W$ is the gradient of the kernel function with respect to
$\ten{x}_i$. If $\Gamma_i = 1$ or $\nabla \Gamma_i =0$, then the particle
distribution may be considered uniform. For example, for particles placed on
an infinite Cartesian mesh $\Gamma_i\approx 1, \nabla \Gamma_i \approx 0$ for
all particles. In this paper, particle distributions that satisfy this
requirement are referred to as uniform, homogeneous, or regular distributions
interchangeably. The term ``particle packing'' is used in the literature as
the process of generating such a homogeneous distribution of particles.

There are many higher-order SPH schemes like
\cite{dilts1999moving,dilts2000moving,bonet_lok:cmame:1999,crksph:jcp:2017}
which employ corrected kernel functions such that $\Gamma_i = 1$ and $\nabla
\Gamma_i = 0$ by construction regardless of the underlying particle
distribution. In such cases it may be argued that there is no requirement for
a uniform/homogeneous distribution of particles. However, these methods do not
always ensure the conservation of linear and angular momentum. On the other
hand, the uncorrected kernels when used with a suitable formulation do
conserve linear and angular momentum as discussed in \cite{crksph:jcp:2017}.
Therefore, even with a higher-order SPH method, it is usually desirable to have
a reasonably uniform distribution of fluid and solid particles to improve
conservation \cite{kiara_sph_2013}.

Many problems involving complex geometries have been solved in the SPH
literature. For example, the flow around blades of mixing devices
\cite{eitzlmayr2014novel} and industrial automotive problems
\cite{chiron2019fast}. In order to simulate such problems, the initial
particle distribution must capture important features of the geometry
accurately while maintaining a uniform spatial distribution.

There have been some earlier attempts to do this in the context of SPH.
\citet{colagrossi2012particle} devised a particle packing algorithm which
uses the kernel gradient to distribute the particles in two-dimensions around
a solid body such that the simulation starts smoothly. In order to generate
the two-dimensional solid, the method proposed by
\citet{marrone-deltasph:cmame:2011} was employed. The method constructs solid
boundaries using piecewise linear curves (PLC). These are translated along the
normal and discretized into equispaced particles according to the desired
particle spacing, $\Delta s$, up to the required number of layers.
\citet{xiao2017new} proposed an algorithm to divide the two-dimensional
domain of interest into square-shaped sub-domains. The sub-domains having an
area equal to the desired area ($\Delta s^2$) are directly converted to
particles. Other particles near the boundary are given partial mass
iteratively. None of these approaches consider three-dimensional geometries.
\citet{dominguez2011development} constructs geometries by clipping the grid
with the boundary of the geometry, and this necessitates using a much higher
resolution to capture the features better. \citet{akinci2013coupling}
proposed a scheme to place particles over triangles of length greater then
the particle spacing. This approach provides a good density distribution and
has been employed to simulate flow for graphics applications in three
dimensions. In a different context, \citet{jiang2015blue} used the SPH method
for packing particles in order to sample blue noise. This method focuses on
packing the particles inside and on the surface of the body. The method does
not generate any particles outside the body. In order to obtain a uniform
particle distribution, a kernel gradient along with a cohesive force proposed
by \citet{akinci2013coupling} is used. It balances the extra force on the
particles near the surface.

In this paper, the methods proposed by \citet{colagrossi2012particle} and
\citet{jiang2015blue} are implemented. The method of \citet{jiang2015blue}
does not by default generate particles in the exterior of the boundary. A
modification of the algorithm is proposed so that it generates the desired
particles in the interior and exterior. A novel SPH-based method to construct
two and three-dimensional geometries, at a given resolution, keeping the
features of the geometry as detailed possible, is proposed. The following
nomenclature is used henceforth to refer to the methods implemented:
\begin{itemize}
\item \emph{Standard}: The geometry is created using the method proposed by
  \citet{marrone-deltasph:cmame:2011} and particles are packed using the
  method proposed by \citet{colagrossi2012particle}. It must be noted that the
  method proposed in \cite{marrone-deltasph:cmame:2011} is limited to
  two-dimensional geometries.
\item \emph{Coupled}: The interior (solid) and exterior (fluid) are created
  separately using the method proposed by \citet{jiang2015blue}. Once these
  converge, the interior and exterior interact using the method of
\cite{colagrossi2012particle}.
\item \emph{Hybrid}: The proposed new method which combines features from the
  above two approaches.
\end{itemize}

The hybrid method uses the kernel gradient to move the particles as proposed
by \citet{colagrossi2012particle} along with a strong repulsion force, which
comes into effect only when two particles are closer than the particle
spacing. The particles on the boundary are allowed to move over the surface
only as done in \cite{jiang2015blue}. Particles that are not on the boundary,
are allowed to move in and out of the boundary surface. Particles are
projected on the boundary which introduces disorder causing the other
particles to adjust accordingly. It is important to note that the proposed
algorithm can be applied in the context of any general-purpose SPH framework.
This makes the approach relatively easy to integrate into SPH codes. The
present implementation uses the open-source PySPH
framework~\cite{pysph2019,PR:pysph:scipy16}. The particle distributions
generated using the above methods are compared for different geometries. The
accuracy of the proposed method is demonstrated by performing an SPH function
and derivative approximation for a known function.

The paper is divided into four sections. The next section briefly discusses
the SPH method. The \cref{sec:alg} describes the algorithms implemented in
detail. In \cref{sec:results}, different geometries are constructed using the
algorithms implemented. In the interest of reproducibility, the implementation
of the algorithms is open source and all the results are fully reproducible.

\section{Smoothed particle hydrodynamics}
\label{sec:sph}

As discussed in the introduction, the SPH method approximates a function using
a smooth, compact kernel, $W(\ten{x})$. Popular choices for the kernel in the
SPH community are the Gaussian \cite{monaghan-review:2005}, splines
\cite{sph:fsf:monaghan-jcp94} and the family of Wendland kernels
\cite{wendland1995piecewise}. In order to reproduce the given function with
$O(h^2)$ accuracy, the kernel function must have
\begin{equation}
  \int_{\Omega} W(\ten{x} - \tilde{\ten{x}}, h) d \tilde{x} = 1
  \text{, and }
  \int_{\Omega} \nabla W(\ten{x} - \tilde{\ten{x}}, h) d \tilde{x} = \ten{0}.
  \label{eq:cond}
\end{equation}
In the continuous approximation, all of these kernel satisfy these properties
\cite{sph:basa-etal-2009}. 
In the SPH method, the domain is discretized using points each
having mass $m$ and density $\rho$. The discrete approximation of the density
$\rho$ and its gradient is given by
\begin{equation}
  \rho_i = \sum_{j=1}^{N} m_j W_{ij}
  \text{ and }
  \nabla \rho_i = \sum_{j=1}^{N} m_j \nabla W_{ij},
  \label{eq:cond1}
\end{equation}
where $W_{ij} = W(x_i-x_j, h)$, $m_j$ is the mass of $i^{th}$ particle and
$N$ is the number of neighbors in the kernel support. We can see that if all
particles have the same density $\rho_i = 1$, then, \cref{eq:cond1} becomes
the same as \cref{eq:gamma} and \cref{eq:del-gamma}. We therefore assume that
all particles have a unit density henceforth. 

For a uniform distribution of particles, the value of $\Gamma_i=1$ and
$\nabla \Gamma_i=0$ accurate to the order $O(h^2)$. Furthermore, as discussed
in \citet{kiara_sph_2013}, the kernel and its gradient can be corrected such
that the condition in \cref{eq:gamma} and \cref{eq:del-gamma} are satisfied
even for a non-uniform distribution of particles. However, using a corrected
kernel introduces issues in momentum conservation when the particles are
non-uniform. Therefore, it is important to have a uniform particle
distribution.

In the weakly compressible SPH scheme, the dynamics of the fluid flow is
governed by
\begin{align}
  \label{eq:ce}
  \frac{d \rho}{d t} & = - \rho \nabla \cdot \ten{u}, \\ \label{eq:mom}
  \frac{d \ten{u}}{d t} & = - \frac{\nabla p}{\rho}  + \nu \nabla^2 \ten{u},\\
  p & = p(\rho, \rho_0, c_0) \label{eq:eos}
\end{align}
where $\rho$, $\nu$, $p$, $\ten{u}$ are the density, kinematic viscosity,
pressure and velocity of a discrete fluid particle respectively. $\rho_0$ and
$c_0$ are the reference density and speed of sound. The derivatives can be
approximated using various forms cf.~\cite{violeau_book_2012}. In a
simulation with a solid body, for example the flow past a cylinder, the solid
particles are represented using classical dummy particles \cite{Adami2012}.
The use of a rectangular lattice would comply with the condition in
\cref{eq:cond1} however, the resulting boundary will be jagged. In order to
produce a body conforming initial particle distribution such that the
conditions in \cref{eq:cond} is satisfied, various particle packing algorithms
are discussed in detail in the next section. We note that once the packing
algorithm completes that there may be slight variations in the density of the
order of around 2-3\%. \citet{colagrossi2012particle} propose that the
particle masses be changed so as to produce a constant density given the
initial pressure field. This may be performed after the packing algorithm is
complete if desired.

\section{Particle packing algorithms}
\label{sec:alg}

In this section, the proposed hybrid algorithm followed by other
algorithms are discussed in detail.

\subsection{Hybrid Algorithm}
\label{sec:hybrid}

\begin{figure}[ht!]
  \centering
  \resizebox{14cm}{7cm}{
  \begin{tikzpicture}
    \draw[line width=0.1mm](0,0) ellipse (2.1cm and 2.1cm);
    \draw[line width=1mm, dash dot](0,0) ellipse (2.1cm and 2.1cm);
    \foreach \x in {0,...,6}
        \foreach \y in  {0,...,3}
            {
              \pgfmathsetmacro{\xc}{\x - 3 * cos(\x / 10)};
              \pgfmathsetmacro{\yc}{2*(\y - 1.5 * cos(\y/10))};
              % Check if numbers are inside circle
              \pgfmathparse{ifthenelse(abs((\xc)^2 + (\yc)^2 - 2.1^2) <= 1.0,%
                  "red",
                  "blue")}
              \fill[\pgfmathresult] (\xc,\yc) circle (0.2);
    };
    \foreach \x in {0,...,6}
        \foreach \y in  {0,...,2}
            {
              \pgfmathsetmacro{\xc}{(\x - 3.0) + 0.5 * cos((\x-3.0)*30)};
              \pgfmathsetmacro{\yc}{2*(\y - 1.5) + 1.0* cos((\y-1.5)*20)};
              % Check if numbers are inside circle
              \pgfmathsetmacro{\res}{ifthenelse(abs((\xc)^2 + (\yc)^2 - 2.1^2) <= 1.0,%
                  "red",
                  "blue")}
              \fill[\res] (\xc,\yc) circle (0.2);
    };
    % external boundary

    \foreach \x in {1,...,9}
        \foreach \y in  {0,5}
            {
              \pgfmathsetmacro{\xc}{\x - 5};
              \pgfmathsetmacro{\yc}{1.5*(\y - 2.5)};
              % Check if numbers are inside circle
              \fill[green] (\xc,\yc) circle (0.2);
    };
    \foreach \x in {1, 9}
        \foreach \y in  {0,...,5}
            {
              \pgfmathsetmacro{\xc}{(\x - 5.0)};
              \pgfmathsetmacro{\yc}{(\y - 2.5)};
              % Check if numbers are inside circle
              \fill[green] (\xc,\yc) circle (0.2);
    };

    \draw[<-] (2.0,0.5) -- node[xshift=2.6cm, yshift=-0.25cm]{Boundary surface} (4,0.0);
    \draw[<-] (1.5,-1.5) -- node[xshift=2.7cm, yshift=-0.25cm]{Boundary node}(4, -2.);
    \draw[<-] (-2,-1.0) -- node[xshift=-2cm, yshift=-1.1cm]{Boundary particle}(-5,-2.5);
    \draw[<-] (3,2) -- node[xshift=2.1cm, yshift=0cm]{Free particle}(4.5,2);
    \draw[<-] (-4,0.5) -- node[xshift=-1.9cm, yshift=0cm]{Frozen particle}(-5,0.5);
  \end{tikzpicture}
  }
  \caption{Schematic of the initial distribution of particles and the
    different kinds of particles.}
  \label{fig:initial}
\end{figure}
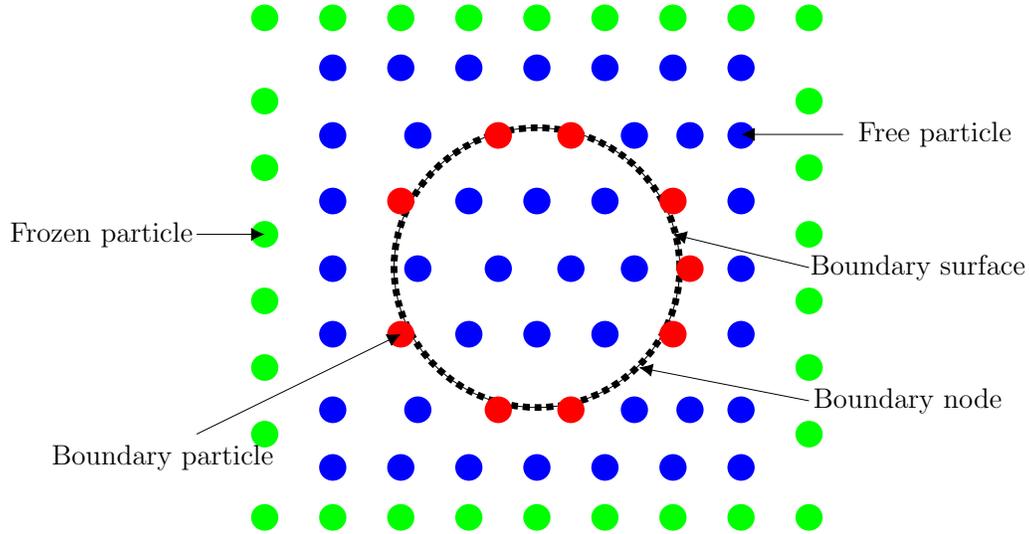

The schematic shown in \cref{fig:initial}, depicts the different kinds of
particles used in the proposed algorithm. In this figure, we consider the case
of a cylinder surrounded by fluid. The dashed black line represents the
surface of the cylinder (boundary surface) which we wish to capture
accurately. We assume that this surface is discretized into a set of points
called ``boundary nodes''. The different kinds of entities shown in the figure
are,
\begin{itemize}
\item \emph{Free particles}: These are particles arranged initially in a
  rectangular or hexagonal-packed pattern. Their motion is not constrained.
  These are depicted as blue circles.
\item \emph{Frozen particles}: These are a set of fixed classical dummy
  particles which surround the free particles in order to provide support to
  the kernel.  These are depicted as green circles.
\item \emph{Boundary particles}: These particles are constrained to move along
  the ``boundary surface'' and are depicted as red circles.
\item \emph{Boundary surface}: The surface of the geometry that is
  discretized. It is represented by a set of fixed points, called ``boundary
  nodes'' which do not influence any other particles.
\item \emph{Boundary node}: These are points that discretize the boundary
  surface, they also store the local surface normals of the boundary surface.
  These are depicted as black dashes. We note that these are called nodes
  because they do not exert any forces on particles and only serve to provide
  information on the position and orientation of the boundary surface.
\end{itemize}
During the algorithm, free particles (blue) may be converted to boundary
particles (red) if they are close to the boundary surface. Once the proposed
algorithm completes, the red boundary particles must conform to the boundary
represented by the dashed black line. The blue particles inside the dashed
boundary will be considered as solid (dummy) particles and those outside as
fluid particles.

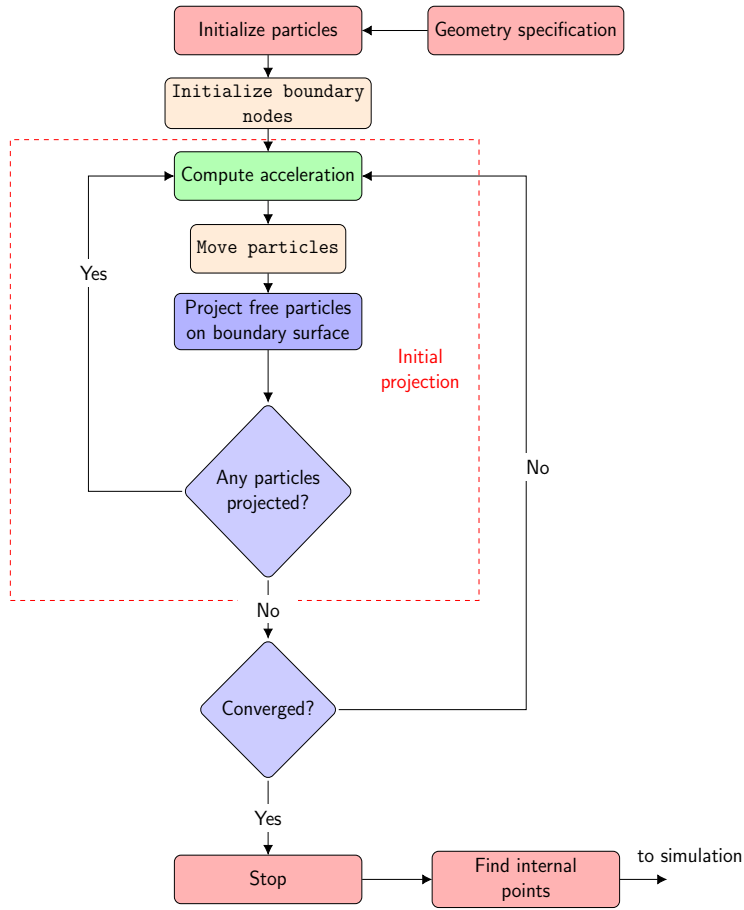
\begin{figure}[ht!]
  \centering
  \resizebox{10cm}{12cm}
  {
  \begin{tikzpicture}[node distance=1.5cm,
    every node/.style={fill=white, font=\sffamily}, align=center]
  % Specification of nodes (position, etc.)
  \node [draw=red, dashed, shape=rectangle, minimum width=10cm, minimum height=9.5cm, anchor=center] at (-0.5,-7) {};
  \node(stage1)[text=red] at (3.25, -7) {Initial \\projection};
  \node (start) [startstop] {Initialize particles};
  \node (initialize) [process, below of=start] {Initialize boundary\\ nodes};
  \node (mesh) [startstop, right of=start, xshift=4cm] {Geometry specification};
  \node (compute) [activityRuns, below of=initialize] {Compute acceleration} ;
  \node (integrate) [process, below of=compute] {Move particles};
  \node (check) [activityStarts, below of=integrate] {Project free particles\\ on boundary surface};
  \node (projcheck) [decision, below of=check, yshift=-2cm]{Any particles\\ projected?};
  \node (convergence) [decision, below of=projcheck, yshift=-3cm] {Converged?};
  \node (stop) [startstop, below of=convergence, yshift=-2cm] {Stop};
  \node (internal) [startstop, right of=stop, xshift=4cm] {Find internal \\ points};

  % Specification of lines between nodes specified above
  % with aditional nodes for description
  \draw[->] (mesh) -- (start);
  \draw[->] (start) -- (initialize);
  \draw[->] (initialize) -- (compute);
  \draw[->] (compute) -- (integrate);
  \draw[->] (integrate) -- (check);
  \draw[->] (check) -- (projcheck);
  \draw[->] (projcheck) -- node[text width=1cm]{No}(convergence);
  \draw[->] (projcheck.west) -- ++(-2,0) -- ++(0,6.5) --
     node[xshift=-0.8cm,yshift=-2cm, text width=1cm]
     {Yes}(compute.west);
  \draw[->] (convergence) --  node[text width=1cm]{Yes}(stop);
  \draw[->] (convergence.east) -- ++(4,0) -- ++(0,11.0) --
     node[xshift=2.0cm, yshift=-6cm, text width=1cm]
     {No}(compute.east);
  \draw[->] (stop) --(internal);
  \draw[->] (internal.east) -- node[xshift=1cm, yshift=0.5cm]{to simulation}++(1.0,0);
  \end{tikzpicture}
}
  \caption{Flowchart of the particle packing algorithm.  The box outlined in
    dashed red lines is the initial projection phase.}
  \label{fig:flowchart}
\end{figure}

The overall flow of the algorithm is shown in \cref{fig:flowchart}. The
algorithm requires two inputs, the geometry information, and the desired
particle spacing. These are to be provided by the user. Given the geometry
surface, we first estimate the number of particles that should lie on this
surface, $N_s$, using the desired spacing of particles and either the length
in 2D or the surface area in 3D.

Initially, the frozen particles are created on the periphery of the domain as
shown in \cref{fig:initial} and placed on a rectangular lattice. Free
particles are then placed inside this on a regular lattice. At this stage, no
boundary particles are identified. The boundary nodes are initialized using
the information provided by the user.

The acceleration on the free particles and any boundary particles (that are
identified later) is computed using a local density gradient, and a repulsive
force. This corresponds to the green block in \cref{fig:flowchart}. The
particles are moved using the computed accelerations. As the free particles
move, they are converted to boundary particles if they are close enough to the
boundary surface. They are then projected to the nearest point on the boundary
surface. These boundary particles are constrained to move only along the
boundary surface. The free particles are iteratively converted to boundary
particles until no free particle is sufficiently close to the boundary.

During the initial projection phase (denoted by the red dashed line in the
\cref{fig:flowchart}), the particles are regularly projected onto the boundary
surface. This is done until the number of boundary particles have reached
$N_s$ and remains there for a few consecutive iterations. The algorithm then
proceeds to settle the particles into a uniform distribution until the
displacement of the particles is less than a user-defined tolerance. This is
denoted as the ``Converged'' block in \cref{fig:flowchart}. Once convergence
is attained, the boundary and free particles inside and outside the surface
are packed as desired. Since the boundary surface is known, free particles can
be easily identified as solid and fluid particles.

\begin{figure}[ht!]
  \centering
  \resizebox{10cm}{7cm}{
  \begin{tikzpicture}
    \draw[line width=0.1mm](0,0) ellipse (2.1cm and 2.1cm);
    \draw[line width=1mm, dash dot](0,0) ellipse (2.1cm and 2.1cm);
    \foreach \x in {0,...,6}
        \foreach \y in  {0,...,3}
            {
              \pgfmathsetmacro{\xc}{\x - 3 * cos(\x / 10)};
              \pgfmathsetmacro{\yc}{2*(\y - 1.5 * cos(\y/10))};
              % Check if numbers are inside circle
              \pgfmathparse{ifthenelse((\xc)^2 + (\yc)^2 <= 2.4^2,%
                  "red",
                  "blue")}
              \fill[\pgfmathresult] (\xc,\yc) circle (0.2);
    };
    \foreach \x in {0,...,6}
        \foreach \y in  {0,...,2}
            {
              \pgfmathsetmacro{\xc}{(\x - 3.0) + 0.5 * cos((\x-3.0)*30)};
              \pgfmathsetmacro{\yc}{2*(\y - 1.5) + 1.0* cos((\y-1.5)*20)};
              % Check if numbers are inside circle
              \pgfmathsetmacro{\res}{ifthenelse((\xc)^2 + (\yc)^2 <= 2.4^2,%
                  "red",
                  "blue")}
              \fill[\res] (\xc,\yc) circle (0.2);
    };
    % external boundary

    \foreach \x in {1,...,9}
        \foreach \y in  {-7,...,0}
            {
              \pgfmathsetmacro{\xc}{\x - 5};
              \pgfmathsetmacro{\yc}{(\y - 4.0)};
              % Check if numbers are inside circle
              \fill[green] (\xc,\yc) circle (0.2);
    };
    \foreach \x in {1,...,9}
        \foreach \y in  {5,...,11}
            {
              \pgfmathsetmacro{\xc}{\x - 5};
              \pgfmathsetmacro{\yc}{(\y - 1.0)};
              % Check if numbers are inside circle
              \fill[green] (\xc,\yc) circle (0.2);
    };
    \foreach \x in {-5,...,1}
        \foreach \y in  {-10,...,11}
            {
              \pgfmathsetmacro{\xc}{(\x - 5.0)};
              \pgfmathsetmacro{\yc}{(\y - 1.0)};
              % Check if numbers are inside circle
              \fill[green] (\xc,\yc) circle (0.2);
    };
    \foreach \x in {9,...,14}
        \foreach \y in  {-10,...,11}
            {
              \pgfmathsetmacro{\xc}{(\x - 5.0)};
              \pgfmathsetmacro{\yc}{(\y - 1.0)};
              % Check if numbers are inside circle
              \fill[green] (\xc,\yc) circle (0.2);
    };

    % rectangle
    \draw[-, dashed, very thick] (-3, -3) -- (3, -3);
    \draw[-, dashed, very thick] (3, -3) -- (3, 3);
    \draw[-, dashed, very thick] (3, 3) -- (-3, 3);
    \draw[-, dashed, very thick] (-3, 3) -- (-3, -3);

    \draw[-, very thick] (-10.5, -11.5) -- (9.5, -11.5);
    \draw[-, very thick] (9.5, -11.5) -- (9.5, 10.5);
    \draw[-, very thick] (9.5, 10.5) -- (-10.5, 10.5);
    \draw[-, very thick] (-10.5, 10.5) -- (-10.5, -11.5);

    \draw[-, very thick] (-8.5, 10.5) -- (-8.5, -11.5);
    \draw[-, very thick] (7.5, 10.5) -- (7.5, -11.5);

    \draw [->] (-11, 0) -- node[xshift=-1.5cm]{\Huge Inlet} (-10, 0);
    \draw [->] (10, 0) -- node[xshift=1.75cm]{\Huge Outlet} (9, 0);

  \end{tikzpicture}
  }
	\caption{The preprocessed patch (dashed) of particles placed in the
    appropriate location of a typical simulation. The blue particles denote
    the free particles identified as fluid particles, the red represents the
    solid particles identified by the packing process. The green particles are
    generated from a fixed mesh of points.}
	\label{fig:sim}
\end{figure}
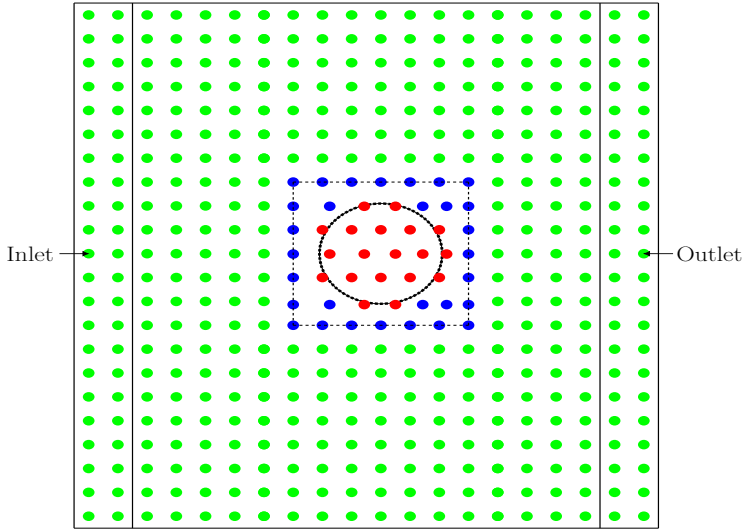

This packed collection of free and boundary particles may be placed into a
larger regular mesh of particles for a simulation. For example, see
\cref{fig:sim} where the dashed region is where the packed particles could be
placed. The exterior of this region (shown in green) can be represented as a
regular mesh of the same spacing. This approach is convenient to use in the
context of fluid flow past solid bodies as done for internal
flows~\cite{negi2020improved} and free surface
flows~\cite{open_bc:tafuni:cmame:2018}. The algorithm described above is
explained in detail in the subsequent sections.

\subsubsection{Initialization of boundary nodes}
\label{subsec:geom}

The boundary nodes which represents the boundary surface are first
initialized. In a two-dimensional domain, a set of points are required which
discretize the boundary curve. The boundary curve may be parametrized by
$\lambda$, and points on the curve may be specified as $\ten{x} = C(\lambda)$
and $ \lambda \in [0, 1]$. This curve is discretized such that $\ten{x}_i =
C(\lambda_i)$. The spacing between points must be such that $|\ten{x}_{i+1}
- \ten{x}_i))| < \Delta s$. The boundary node coordinates are initialized
using these points. The outward normals $n_{x},n_{y}$ for any node $i$ are
calculated using
\begin{equation}
  \begin{split}
  n_{x,i}  & =0.5  \left ( \frac{y_{i+1} - y_{i}}{d_{i+1, i}} +
   \frac{y_{i} - y_{i-1}}{d_{i, i-1}} \right) \\
  n_{y,i}  & =-0.5 \left( \frac{x_{i+1} - x_{i}}{d_{i+1, i}} +
  \frac{x_{i} - x_{i-1}}{d_{i, i-1}} \right)
  \end{split}
  \label{eq:2dnormal}
\end{equation}
where $d_{i, j}$ is the length of the segment joining node at $(x_i, y_i)$ and
$(x_j, y_j)$. The resulting normal is then normalized. The \cref{eq:2dnormal}
ensures that sharp corners of the curve have smooth normals. For a
three-dimensional case, a triangulation of the surface with outward normals is
necessary. The centroid of each triangle and its normal is used to initialize
the boundary nodes.

In SPH, the actual boundary surface is exactly in between solid and fluid
particles. Thus, both in two and three dimensions, given a particle spacing of
$\Delta s$, the boundary nodes are shifted by $\Delta s / 2$ inside the actual
boundary to correctly implement the solid boundary conditions as discussed in
\citet{marrone-deltasph:cmame:2011}. In order to move the nodes inwards, the
following translation is performed on each boundary node given by,
\begin{equation}
  \label{eq:trans}
  \ten{x} = \ten{x} - \frac{\Delta s}{2} \hat{\ten{n}},
\end{equation}
where, $\ten{x}$ is the position of the node and $\hat{\ten{n}}$ is its unit
normal pointing outwards. It must be noted that this is optional and one can
provide a pre-shifted surface and avoid \cref{eq:trans}.

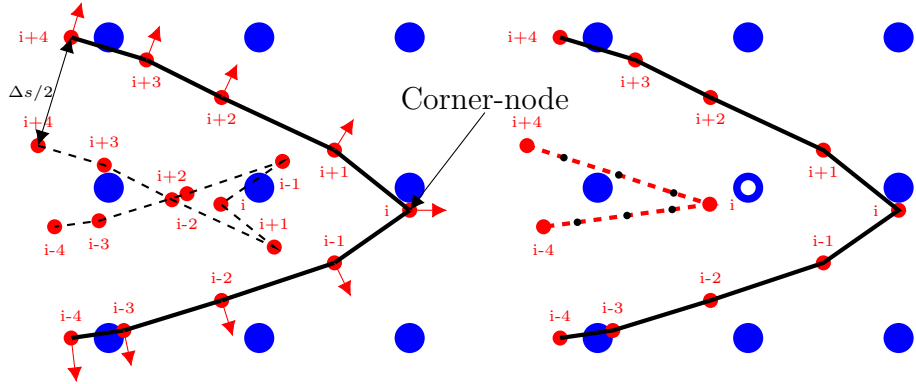
\begin{figure}[h!]
  \centering
  \begin{tikzpicture}
    \foreach \x in {2,...,4}
        \foreach \y in  {0,...,2}
            {
              \pgfmathsetmacro{\xc}{\x*2};
              \pgfmathsetmacro{\yc}{\y*2};
              % Check if numbers are inside circle
              \fill[blue] (\xc,\yc) circle (0.2);
    };

    \foreach \x in {2,...,4}
        \foreach \y in  {0,...,2}
            {
              \pgfmathsetmacro{\xc}{\x*2 + 6.5};
              \pgfmathsetmacro{\yc}{\y*2};
              % Check if numbers are inside circle
              \fill[blue] (\xc,\yc) circle (0.2);
    };

    \fill[red] (3.5, 4) circle(0.1) node[xshift=-0.5cm]{\tiny i+4};
    \fill[red] (4.5, 3.7) circle(0.1)node[yshift=-0.3cm]{\tiny i+3};
    \fill[red] (5.5, 3.2) circle(0.1)node[yshift=-0.3cm]{\tiny i+2};
    \fill[red] (7, 2.5) circle(0.1)node[yshift=-0.3cm]{\tiny i+1};
    \fill[red] (8, 1.7) circle(0.1)node[xshift=-0.3cm]{\tiny i};
    \fill[red] (7, 1.) circle(0.1)node[yshift=0.3cm]{\tiny i-1};
    \fill[red] (5.5, 0.5) circle(0.1)node[yshift=0.3cm]{\tiny i-2};
    \fill[red] (4.2, 0.1) circle(0.1)node[yshift=0.3cm]{\tiny i-3};
    \fill[red] (3.5, 0) circle(0.1)node[yshift=0.3cm]{\tiny i-4};

    \draw[-, ultra thick] (3.5, 4) -- (4.5, 3.7) --(5.5, 3.2)-- (7, 2.5) -- (8, 1.7) -- (7, 1.) -- (5.5, 0.5) -- (4.2, 0.1)-- (3.5, 0);

    \fill[red] (3.5+6.5, 4) circle(0.1) node[xshift=-0.5cm]{\tiny i+4};
    \fill[red] (4.5+6.5, 3.7) circle(0.1)node[yshift=-0.3cm]{\tiny i+3};
    \fill[red] (5.5+6.5, 3.2) circle(0.1)node[yshift=-0.3cm]{\tiny i+2};
    \fill[red] (7+6.5, 2.5) circle(0.1)node[yshift=-0.3cm]{\tiny i+1};
    \fill[red] (8+6.5, 1.7) circle(0.1)node[xshift=-0.3cm]{\tiny i};
    \fill[red] (7+6.5, 1.) circle(0.1)node[yshift=0.3cm]{\tiny i-1};
    \fill[red] (5.5+6.5, 0.5) circle(0.1)node[yshift=0.3cm]{\tiny i-2};
    \fill[red] (4.2+6.5, 0.1) circle(0.1)node[yshift=0.3cm]{\tiny i-3};
    \fill[red] (3.5+6.5, 0) circle(0.1)node[yshift=0.3cm]{\tiny i-4};

    \draw[-, ultra thick] (3.5+6.5, 4) -- (4.5+6.5, 3.7) --(5.5+6.5, 3.2)-- (7+6.5, 2.5) -- (8+6.5, 1.7) -- (7+6.5, 1.) -- (5.5+6.5, 0.5) -- (4.2+6.5, 0.1)-- (3.5+6.5, 0);

    \draw[red, ->] (3.5, 4) -- (3.64, 4.47);
    \draw[red, ->] (4.5, 3.7) -- (4.68, 4.16);
    \draw[red, ->] (5.5, 3.2) -- (5.71, 3.65);
    \draw[red, ->] (7, 2.5) -- (7.26, 2.92);
    \draw[red, ->] (8, 1.7) -- (8.5, 1.7);
    \draw[red, ->] (7, 1.) -- (7.22, 0.55);
    \draw[red, ->] (5.5, 0.5) -- (5.65, 0.02);
    \draw[red, ->] (4.2, 0.1) -- (4.30, -0.38);
    \draw[red, ->] (3.5, 0) -- (3.57, -0.59);

    \fill[red] (3.06, 2.56) circle(0.1) node[yshift=0.3cm] {\tiny i+4};
    \fill[red] (3.94, 2.30) circle(0.1) node[yshift=0.3cm] {\tiny i+3};
    \fill[red] (4.84, 1.84) circle(0.1) node[yshift=0.3cm] {\tiny i+2};
    \fill[red] (6.20, 1.21) circle(0.1) node[yshift=0.3cm] {\tiny i+1};
    \fill[red] (5.49, 1.78) circle(0.1) node[xshift=0.3cm] {\tiny i};
    \fill[red] (6.31, 2.35) circle(0.1) node[yshift=-0.3cm, xshift=0.1cm] {\tiny i-1};
    \fill[red] (5.04, 1.92) circle(0.1) node[yshift=-0.4cm] {\tiny i-2};
    \fill[red] (3.87, 1.56) circle(0.1) node[yshift=-0.3cm] {\tiny i-3};
    \fill[red] (3.28, 1.48) circle(0.1) node[yshift=-0.3cm] {\tiny i-4};

    \draw[dashed, -, thick](3.06, 2.56) -- (3.94, 2.30) -- (4.84, 1.84) --
    (6.20, 1.21) -- (5.49, 1.78) -- (6.31, 2.35) -- (5.04, 1.92) -- (3.87,
    1.56) -- (3.28, 1.48);

    \draw[red, -, dashed, ultra thick](3.06+6.5,2.56) -- (3.55+6.5,2.40) --
    (4.28+6.5,2.17) -- (5.01+6.5,1.93) -- (5.49+6.5,1.78) -- (4.99+6.5,1.71)
    -- (4.39+6.5,1.63) -- (3.73+6.5,1.54) -- (3.28+6.5,1.48);

    \fill[red] (3.06+6.5, 2.56) circle(0.1) node[yshift=0.3cm] {\tiny i+4};
    \fill[black] (3.55+6.5,2.40) circle(0.05);
    \fill[black] (4.28+6.5,2.17) circle(0.05);
    \fill[black] (5.01+6.5,1.93) circle(0.05);
    \fill[red] (5.49+6.5, 1.78) circle(0.1) node[xshift=0.3cm] {\tiny i};
    \fill[black] (4.99+6.5,1.71) circle(0.05);
    \fill[black] (4.39+6.5,1.63) circle(0.05);
    \fill[black] (3.73+6.5,1.54) circle(0.05);
    \fill[red] (3.28+6.5, 1.48) circle(0.1) node[yshift=-0.3cm] {\tiny i-4};

    \draw[<-] (8, 1.7) -- (9, 3.0) node[yshift=0.2cm]{Corner-node};
    \draw[<->] (3.5, 4) -- (3.06, 2.56) node[yshift=0.7cm, xshift=-0.1cm]{\tiny $\Delta s/2$};
    \fill[white] (6+6.5,2) circle (0.1);
\end{tikzpicture}
\caption{Shifting of the boundary near a sharp edged boundary. The boundary
  nodes are depicted in red with normals. On the left is the boundary surface
  after the initial shifting shown as black dashed lines. On the right is the
  final geometry after removal of the intersecting edges which are shown as
  dashed red lines. The annular blue free particle is the candidate to be
  placed on the corner and held fixed.}
	\label{fig:corner_node}
\end{figure}

Care must be taken when there are sharp changes in the features of the
geometry. Consider an airfoil trailing edge in \cref{fig:corner_node} shown as
a black line. These sharp corners are marked as ``corner nodes'', and in this
case, it is the $i$\textsuperscript{th} node. When the nodes on this surface
are shifted, the boundary surface tends to self-intersect itself. This is
shown by the red nodes connected using a black dashed line. In order to remove
the intersection of the boundary surface near the corner node, the points
$i-3$ to $i-1$ are replaced by the points on the line joining $i-4$ and $i$
with equal spacing shown by black points. Similarly the points $i+1$ to $i+3$
are also replaced to lie along the line joining points $i$ and $i+4$. This
results in a non-intersecting surface as shown by the red dashed line in the
right side of \cref{fig:corner_node}. Once the intersection is resolved, the
nearest free particle near the corner node (annular blue particle) is placed
on it and converted to a \emph{fixed boundary particle}. The position of these
fixed boundary particles do not change in the entire simulation.

In the case of a three-dimensional object, one has to make sure that the
surface does not intersect after applying \cref{eq:trans} or use a pre-shifted
surface as an input.

\subsubsection{Dynamics of the particles}
\label{subsec:pm}

In this section, the dynamics of particle regularization is discussed.
Two forces are applied on the particles and together these regularize the
particle distribution. The two forces are, a gradient due to particle
disorder and a pure inter-particle repulsive force.

In the presence of a constant pressure field, $p_b$ and no viscous effect, the
momentum equation (\cref{eq:mom}), becomes
\begin{equation}
  \label{eq:backgrounpressure}
  \frac{d \ten{u}}{dt} = -\frac{\nabla (1 \cdot p_b)}{\rho} =
  -\frac{p_b \nabla(1) + \nabla(p_b)}{\rho}.
\end{equation}

When the term, $p_b \nabla(1)$, on the right hand side is discretized using
the SPH method, we obtain $p_b \nabla \Gamma$ (see \cref{eq:del-gamma}). This
is non-zero when the particles are not uniform and hence particles exert a
force on each other in order to reach an equilibrium position. Using the SPH
approximation, the above equation is discretized as
\begin{equation}
  \label{eq:backgrounpressure_sph}
  a_{b,i} = \frac{d \ten{u}_i}{dt} =
  -\sum_j p_b \frac{V_i V_j}{m_i}\nabla W_{ij},
\end{equation}
where $V_i = \frac{m_i}{\rho_i}$ is the volume of the $i$\textsuperscript{th}
particle, $\rho_i$ is the density, $W_{ij}$ is the kernel function chosen for
the SPH discretization. The summation is over all the neighbors of
$i^\text{th}$ particle. Since all SPH kernels satisfy \cref{eq:cond}, any
suitable SPH kernel discussed in \cref{sec:sph} can be employed. In this
paper, the quintic spline kernel is used given by,
\begin{equation}
	\label{eq:quintic-spline}
	W(q) = \left \{
	\begin{array}{ll}
		\sigma \left[ {(3-q)}^5 - 6{(2-q)}^5 + 15{(1-q)}^5 \right] \
		& \textrm{for} \ 0\leq q \leq 1,\\
		\sigma \left[ {(3-q)}^5 - 6{(2-q)}^5 \right]
		& \textrm{for} \ 1 < q \leq 2,\\
		\sigma \ {(3-q)}^5  & \textrm{for} \ 2 < q \leq 3,\\
		0 & \textrm{for} \ q>3,\\
	\end{array} \right.
\end{equation}
where, $\sigma = 1/(120 h), 7/(478\pi h^2),1/(120\pi h^3) $ in one, two and
three-dimensions respectively and $q=|\ten{r}|/h$. In this algorithm, the
value of $p_b$ is set to $1$ independent of the resolution. In addition to
this force, a repulsive force (RF) similar to the gradient of the Lennard
Jones potential (LJP) is used. The new repulsion force potential ($\phi_{RF}$) is
given by
\begin{equation}
  \label{eq:LJP}
  \phi_{RF} = 12 \ k_r \left( \frac{c^2}{r^3} - \frac{c}{r^2} \right)\\
\end{equation}
where $k_r$ is a constant. We set $c = 2 \alpha \Delta s/3 $, where $\alpha$
is a scaling factor. The gradient of \cref{eq:LJP} gives us the force due to
$\phi_{RF}$. The force is kept constant for $r<\Delta s / 2$ in order to avoid
very large repulsion forces. The SPH approximation of the acceleration due to
\cref{eq:LJP} can be written as
\begin{equation}
  \label{eq:LJPacc_sph}
  a_{RF,i} = -\nabla \phi_{RF,i} =
  \left \{
    \begin{array}{ll}
    \sum_j 192 \ k_r \ten{e}_{ij}
    \left( \frac{3  c^2}{\Delta s^4} - \frac{c}{\Delta s^3} \right)& r_{ij} \leq \Delta s/2\\
    \sum_j 12 \ k_r \ten{e}_{ij}
    \left( \frac{3c^2}{r_{ij}^4} - \frac{2c}{r_{ij}^3} \right)
    & \Delta s/2 < r_{ij} \leq \alpha \Delta s \\
    0 & r_{ij} >  \alpha \Delta s\\
    \end{array} \right.
\end{equation}
where $\ten{e}_{ij} = \ten{x}_{ij}/r_{ij}$. Note that the acceleration is
continuous at $r_{ij} = \Delta s/2$. We find that using a value of
$\alpha=0.95$ works well for the algorithm.

\begin{figure}[ht!]
  \centering
  \ifthenelse{\showimages=1}
  {
    \includegraphics[width=\textwidth]{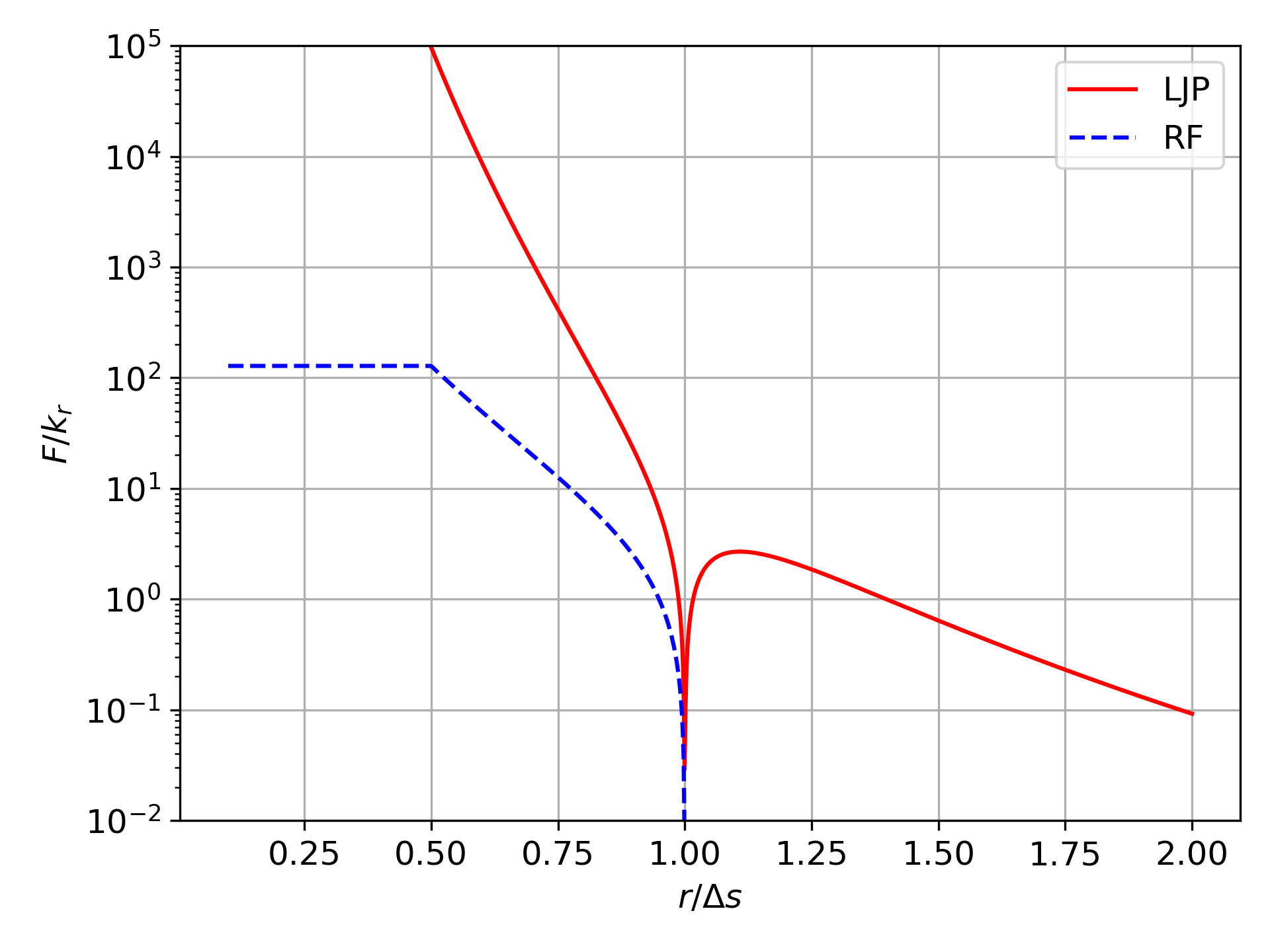}
  }{}
  \caption{Force due to LJ potential and the gradient of \cref{eq:LJP} as a
    function of distance, $r$ and $\alpha=1$.}
  \label{fig:ljp}
\end{figure}

It is clear from \cref{eq:LJPacc_sph} that this force is active only when
particles come closer than the desired particle spacing. This prevents
particle pairing, which may happen due to the use of some kernels like cubic
spline for large time steps \cite{dehnen-aly-paring-instability-mnras-2012}.
In \cref{fig:ljp}, the comparison between the force due to LJP and RF is
shown. The LJP repulsion force increases rapidly compared to our suggested
repulsion force. This allows us to use a larger time step during integration.
Moreover, unlike the force due to LJP, the new force does not introduce
inter-particle attraction.

As discussed in~\cite{colagrossi2012particle}, for stability, a damping force
is used to reduce the energy of the system. The acceleration due to damping
for the $i^\text{th}$ particle is given by
\begin{equation}
  \label{eq:damp_sph}
 a_{d,i} = -\zeta \ten{u}_i
\end{equation}
where $\zeta$ is the damping constant and $\ten{u}_i$ is the velocity of the
$i^\text{th}$. The value of the damping constant $\zeta$ is discussed in
\cref{subsec:time}. Thus, the equation governing the dynamics of the system
is given by
\begin{equation}
  \label{eq:total_force}
 \frac{du}{dt} = -\frac{\nabla p_b}{\rho} - \nabla \phi_{RF} - \zeta \ten{u}
\end{equation}
The above equation can be converted into SPH form using
\cref{eq:backgrounpressure_sph,eq:LJPacc_sph,eq:damp_sph}. It must be noted
that the combination of background pressure force and the repulsion force
produces repulsion only when particles are disordered. This can also be
accomplished by using the particle shifting techniques (PST) first proposed by
\cite{xu2009accuracy}.

On using \cref{eq:total_force} the velocities and new positions are calculated
using a semi-implicit Euler integration given by,
\begin{equation}
  \label{eq:velocity}
  \begin{split}
 u_i(t + \Delta t) &= u_i(t) + \Delta t a_{i}(t) \\
 r_i(t + \Delta t) &= r_i(t) + \Delta t \ u_i(t+\Delta t)
\end{split}
\end{equation}
In the case of boundary particles, the velocities are corrected to constrain
them to move along the surface (discussed in \cref{subsec:om}). It should be
noted that the packing algorithm only ensures that particle distributions are
regular and therefore using a higher order integrator would not promise
better results.

The stability of the method in a two and three-dimensional domain for finite
perturbation under the action of forces described above is studied. A domain
of size 1 unit along each coordinate direction is considered. Particles are
placed with a spacing of $\Delta s = 0.05$. Fixed particles are placed
outside this unit box suitably. The value of $p_{b}$, $\zeta$, $k_r$ and
$\Delta t$ are set as discussed in \cref{subsec:time}. A single particle
close to the center is perturbed by $\Delta s / 2$ in each direction. The
particles are moved using \cref{eq:total_force} and \cref{eq:velocity} for
15000 iterations. The stability of the following initial distribution of
particles is considered:

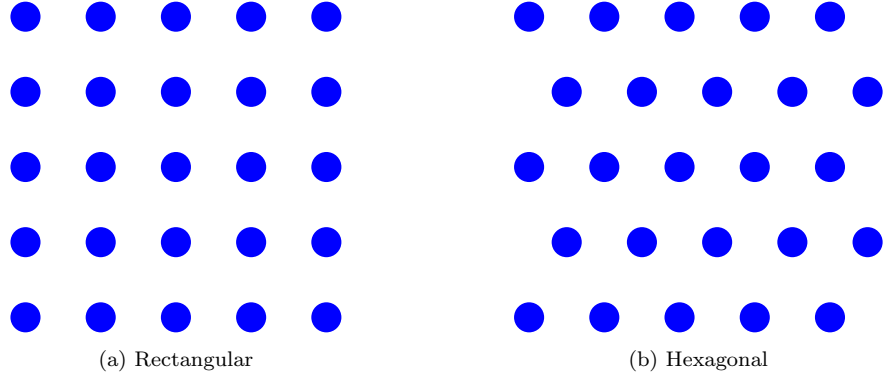
\begin{figure}[ht!]
  \begin{subfigure}{0.5\textwidth}
    \centering
    \begin{tikzpicture}
      \foreach \x in {0,...,4}
          \foreach \y in  {0,...,4}
              {
                \pgfmathsetmacro{\xc}{\x};
                \pgfmathsetmacro{\yc}{\y};
                % Check if numbers are inside circle
                \fill[blue] (\xc,\yc) circle (0.2);
      };
    \end{tikzpicture}
    \caption{Rectangular}
    \label{fig:rect}
  \end{subfigure}
  \begin{subfigure}{0.5\textwidth}
    \centering
    \begin{tikzpicture}
      \foreach \x in {0,...,4}
          \foreach \y in  {0,...,2}
              {
                \pgfmathsetmacro{\xc}{\x};
                \pgfmathsetmacro{\yc}{\y*2.};
                % Check if numbers are inside circle
                \fill[blue] (\xc,\yc) circle (0.2);
      };
      \foreach \x in {0,...,4}
          \foreach \y in  {0,...,1}
              {
                \pgfmathsetmacro{\xc}{\x*1.+0.5};
                \pgfmathsetmacro{\yc}{\y*2.+1};
                % Check if numbers are inside circle
                \fill[blue] (\xc,\yc) circle (0.2);
      };
    \end{tikzpicture}
    \caption{Hexagonal}
    \label{fig:hex}
  \end{subfigure}
  \caption{Different packing structures in 2D.}
  \label{fig:packing}
\end{figure}

\begin{itemize}
  \item hexagonal packing shown in \cref{fig:hex} with number density ($ND$) gradient.
  \item hexagonal packing shown in \cref{fig:hex} with $ND$ + repulsion force ($RF$).
  \item rectangular packing shown in \cref{fig:rect} with $ND$ gradient.
  \item rectangular packing shown in \cref{fig:rect} with $ND + RF$.
\end{itemize}

\begin{figure}[ht!]
  \begin{subfigure}{0.5\textwidth}
    \centering
    \ifthenelse{\showimages=1}
    {
    \includegraphics[width=\textwidth]{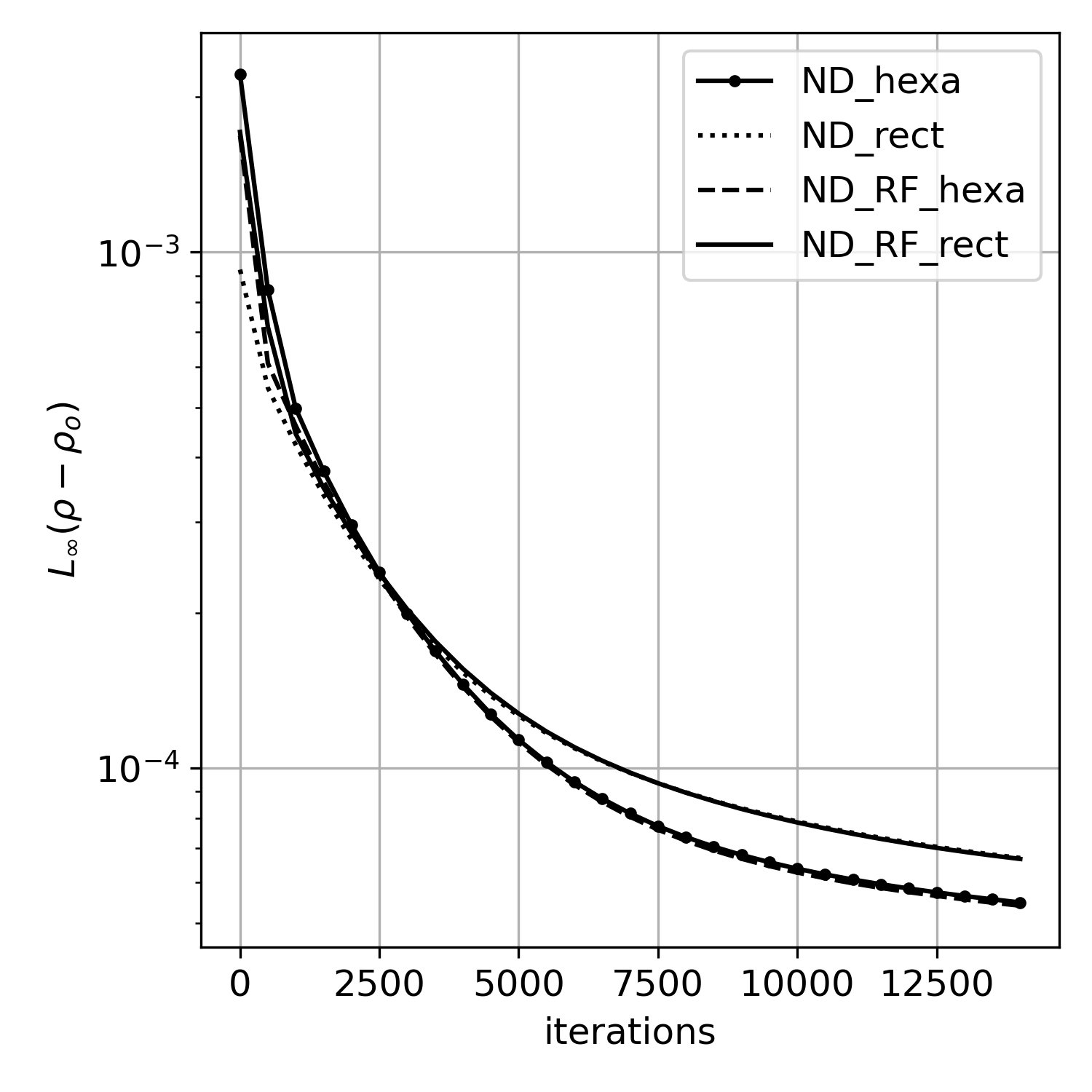}
    }{}
    \caption{2D}
    \label{fig:2d1}
  \end{subfigure}
  \begin{subfigure}{0.5\textwidth}
    \centering
    \ifthenelse{\showimages=1}
    {
    \includegraphics[width=\textwidth]{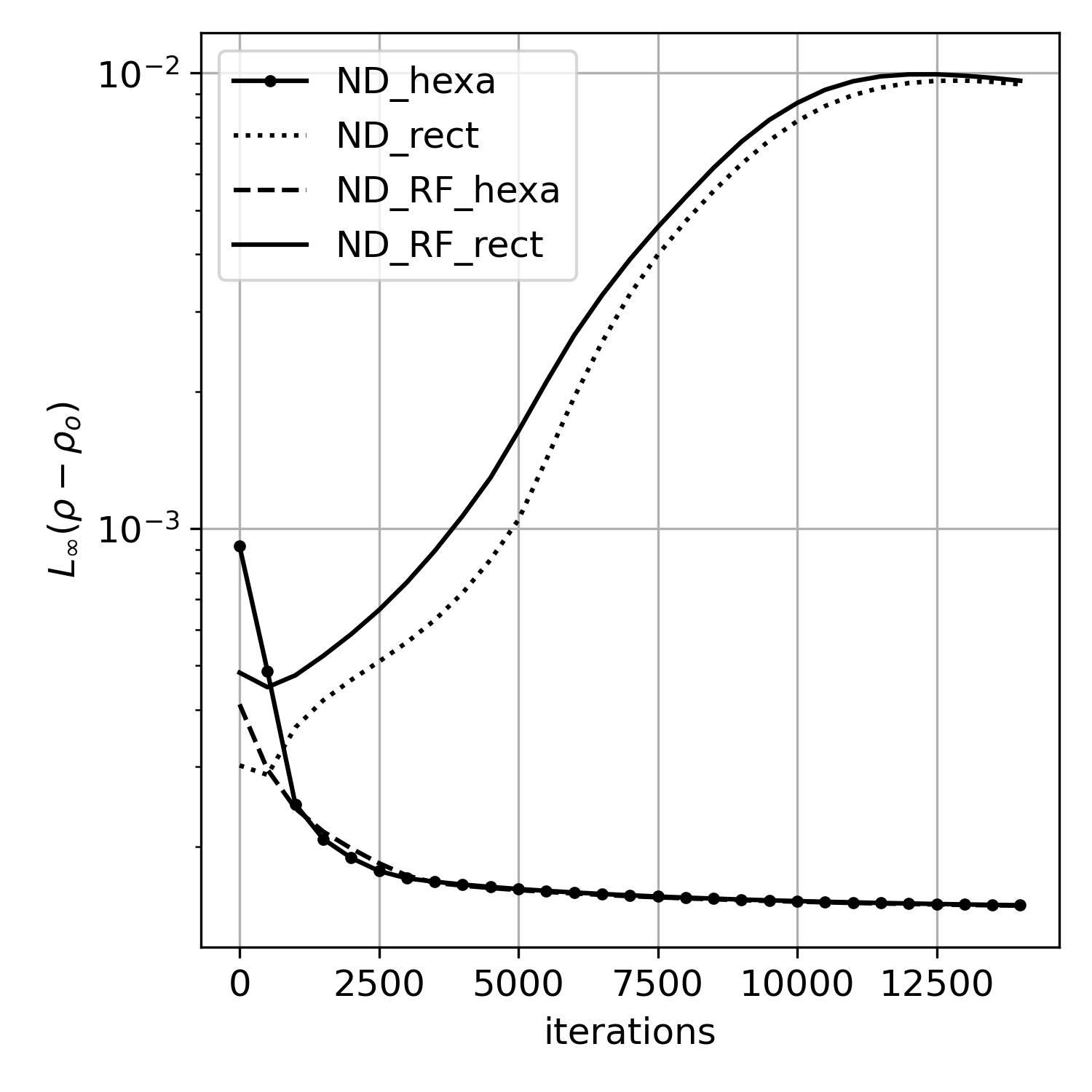}
    }{}
    \caption{3D}
    \label{fig:3d1}
  \end{subfigure}
  \caption{Density convergence with a particle at center perturbed by $\Delta
  s/4$.}
  \label{fig:pert}
\end{figure}

The error in every iteration is evaluated using
\begin{equation}
  L_{\infty}(\rho - \rho_{o}) = \max(\{(\rho_i - \rho_{o}) ,  \forall i \in 1..N \})
\end{equation}
where $N$ is the number of particles in the domain. In \cref{fig:2d1} and
\cref{fig:3d1} the $L_{\infty}(\rho - \rho_{o})$ is plotted with number of
iterations for 2D and 3D domain respectively with $\rho_{o}=1.0$. It is
evident from the figure that in 2D, all the combinations perform well. In the
case of the 3D domain, rectangular lattices settle into an equilibrium
configuration with a much larger density difference. Therefore the
rectangular lattice in 3D is an unstable equilibrium. In contrast, the
hexagonal packing in 3D is in a stable equilibrium.

This behavior in 3D can be understood from the total potential energy of the
particles. It can be seen that \cref{eq:gamma} for $\Gamma_i$ is the
potential energy of the $i$\textsuperscript{th} particle and hence $\sum_i
\Gamma_i $ is the total potential energy of the system. The acceleration of a
particle given in \cref{eq:backgrounpressure} is the negative of the gradient
of its potential energy. We have numerically found that a small perturbation
of a particle at a given site reduces the total potential energy in the case
of the rectangular lattice whereas, the same perturbation results in higher
potential energy in the case of the hexagonal packing. This suggests that the
hexagonal packing is stable in 3D unlike the rectangular lattice. A careful
analysis on the stability is outside the scope of the present work. Due to
the stability of hexagonal packing shown in \cref{fig:hex}, it is used in all
our test cases.

\subsubsection{Projecting free particles to the surface}
\label{subsec:pa_add}

\begin{figure}
  \centering
  \includegraphics[width=0.8\textwidth]{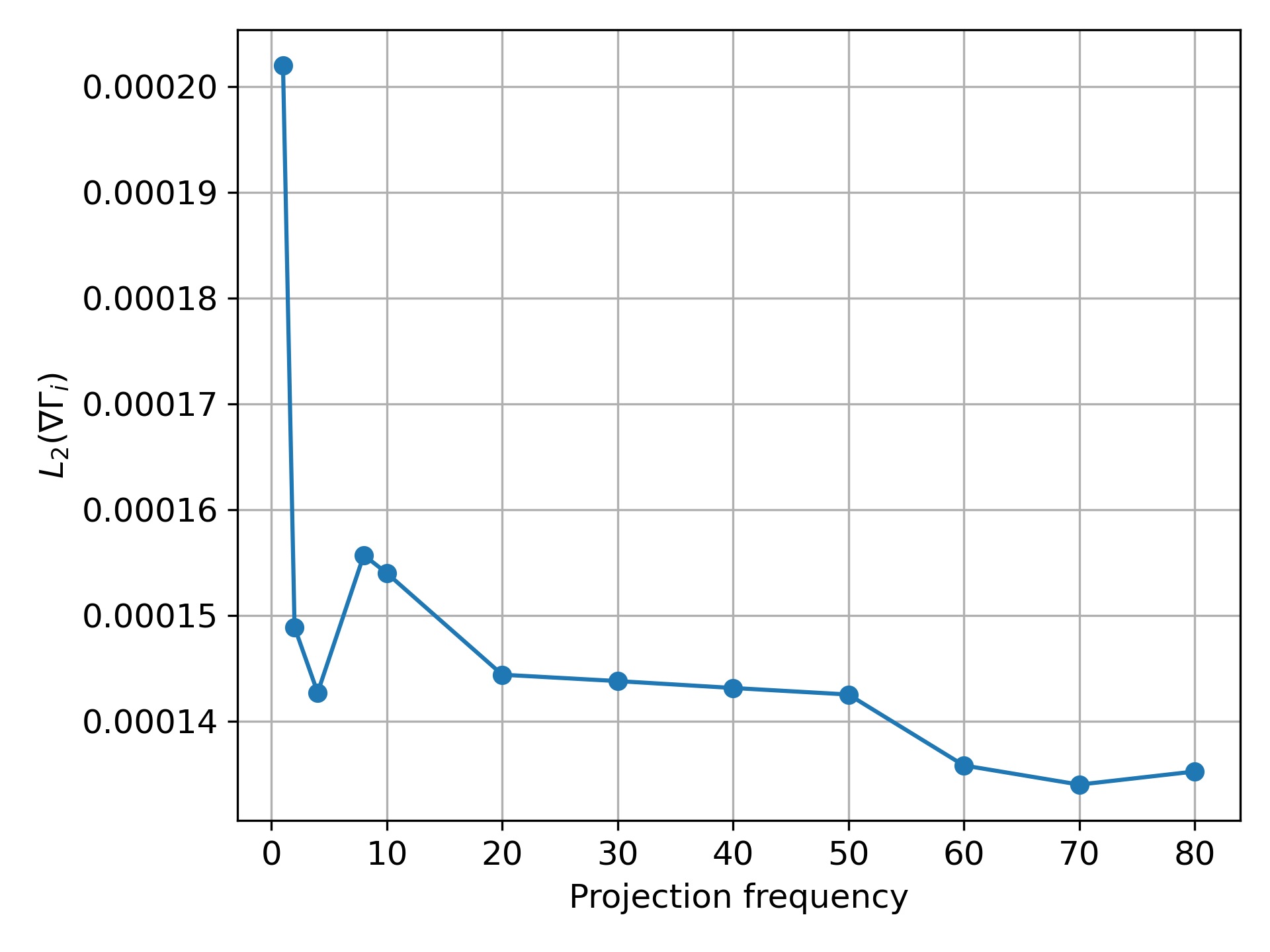}
  \caption{Error in $\nabla \Gamma_i$ for domain with unit radius cylinder with different projection frequency.}
  \label{fig:dfreq}
\end{figure}

Initially, the boundary particles are unlikely to conform to the boundary
surface. At $t=0$ no particles are assigned as boundary particles. Free
particles are converted to boundary particles after every $50$ iterations.
This is called as the ``projection frequency''. The projection frequency is a
user-defined parameter. Increasing this does not guarantee better results
however a very small value is not recommended. On running the proposed
algorithm with spacing $\Delta s = 0.1$ around a unit radius 2D cylinder with
varying projection frequency as shown in \cref{fig:dfreq}, it was found that
the after projection frequency $20$, the number density gradient do not change
by a large value. Thus, the value $50$ is chosen heuristically.

\begin{algorithm}[h!]
  \SetAlgoLined
  \KwIn{\lstinline{proj_freq, ds=}$\Delta s$}
  \KwResult{List of free particles to be projected}
  \code{iteration} = 0\;
  \lstinline{empty_count} = 0\;
  \While{\codek{not converged}} {
    \ldots\;
    \If{\codek{iteration \% proj_freq == 0}}{
      \eIf{ \lstinline{empty_count < 4}} {
        \lstinline{p_list = FindParticlesNearBoundary}(\code{maxdist=0.5*ds)}\;
        \eIf {\codek{len(p_list) > 0}}{
          \code{ConvertParticles()}\;
          \lstinline{empty_count} = 0\;
        }{
          \lstinline{empty_count}++\;
        }

      } {
        \lstinline{p_list = FindParticlesNearBoundary(maxdist=0.65*ds)}\;
        \code{ConvertParticles()}\;
      }
      \code{ProjectToBoundary()}\;
      \ldots\;
    }
    \code{iteration++}\;
  }
  \caption{Pseudo-code for free particle projection.}
  \label{alg:proj}
\end{algorithm}

If we consider a flat surface, an estimate for the number of particles that
can fill the surface is, $N_s= A_s/\Delta s^{(d-1)}$ where $A_s$ is the area
of the surface and $d$ is the dimension of the space in which the surface is
embedded. In order to perform projection, two different criteria are employed
depending upon whether the initial projection in \cref{fig:flowchart} is
complete or not, as shown in algorithm \ref{alg:proj}.

\begin{sloppypar}
In the algorithm, the function \lstinline{FindParticlesNearBoundary} finds all
the free particles that are less than a prescribed distance,
\lstinline{maxdist}, to a boundary node. The distance is computed as follows,
given a free particle $p$, we find the boundary node $b$ that is closest and
compute the distance $\ten{r}_{pb} \cdot \ten{n}_b$, where $\ten{r_{pb}} =
\ten{r}_p - \ten{r}_b$ and $\ten{n}_b$ is the normal at $b$. It is to be noted
that the algorithm to find the nearest boundary node uses the existing
neighbors generated when computing the accelerations discussed in
\cref{subsec:pm}.
\end{sloppypar}

\begin{sloppypar}
The \lstinline{empty_count} variable stores the number of consecutive passes
for which the \lstinline{p_list} was empty. The free particles having a
distance to the boundary node less than $0.5 \Delta s$ are converted to
boundary particles in \code{ConvertParticles} and projected to the boundary
surface in \code{ProjectToBoundary}. However, if the \lstinline{empty_count}
exceeds 3, the distance threshold is increased to $0.65 \Delta s$. This
threshold is increased to handle the corner cases where free particle are
just outside $0.5 \Delta s$ distance\footnote{This value was found based on
numerical experiments. Too small a value like $0.55 \Delta s$ produces poor
results in coarse resolutions and too large a value ($0.75 \Delta s$)
produced poor distributions with finer resolutions.}. This iterative
conversion of free particles to boundary particles is necessary in order to
capture the surface accurately.
\end{sloppypar}

\subsubsection{Kinematics of boundary particles}
\label{subsec:om}

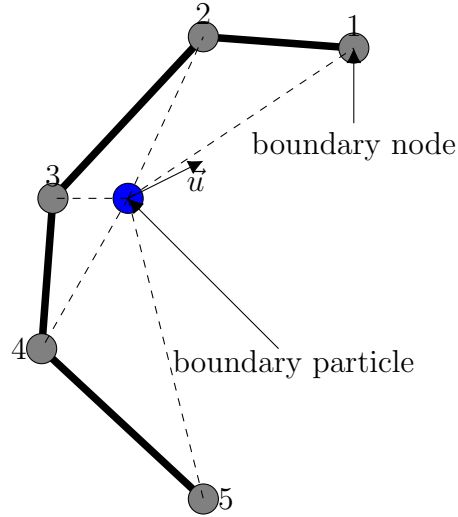
\begin{figure}[ht!]
  \centering

  \begin{tikzpicture}

    \draw [-, line width=1mm] (2,-4) -- (-0.15,-2)-- (0, 0) -- (2,2.15) -- (4, 2);
    \draw[fill=gray](0,0) ellipse (2mm and 2mm) node[yshift=3mm]{3};
    \draw[fill=gray](2,-4) ellipse (2mm and 2mm) node[xshift=3mm]{5};
    \draw[fill=gray](4,2) ellipse (2mm and 2mm) node[yshift=3mm]{1};
    \draw[fill=gray](-0.15,-2) ellipse (2mm and 2mm) node[xshift=-3mm]{4};
    \draw[fill=gray](2,2.15) ellipse (2mm and 2mm) node[yshift=3mm]{2};

    \draw[fill=blue] (1,0) ellipse (2mm and 2mm);
    \draw[->] (1,0) -- node[xshift=0.4cm]{$\vec{u}$}(2,0.5);

    \draw[-, dashed] (1,0) -- (2,-4);
    \draw[-, dashed] (1,0) -- (-0.15,-2);
    \draw[-, dashed] (1,0) -- (0,0);
    \draw[-, dashed] (1,0) -- (2,2.15);
    \draw[-, dashed] (1,0) -- (4,2);

    \draw[<-] (1,0) -- node[xshift=1.2cm, yshift=-1.2cm]{boundary particle}(3,-2);
    \draw[<-] (4,2) -- node[yshift=-0.8cm]{boundary node}(4, 1);

    \end{tikzpicture}
  \caption{Motion of boundary particle along the geometry.}
  \label{fig:bm}
\end{figure}

As discussed earlier, the movement of the boundary particles are constrained
along the boundary surface. \Cref{fig:bm} illustrates the motion of a boundary
particle (in blue) along the geometry represented by nodes 1, 2, 3, 4, and 5.
We note that even though the particle is a boundary particle, it is only
projected onto the surface every \lstinline{proj_freq} time steps as discussed
in \cref{subsec:pa_add} hence the boundary particle may not exactly lie on the
boundary as it moves.

The motion of the boundary particle is performed as follows. The boundary node
nearest to the boundary particles in the direction of the particle's velocity
is identified. Consider a boundary particle $p$, and a boundary node $j$, near
$p$ having position $\ten{x}_{p}$ and $\ten{x}_j$ respectively. The index of
the nearest node, $J$ is determined by
\begin{equation}
  \label{eq:dir}
  J = \argmin_j \{r_{pj}\ |\ j \in \mathcal{N}_{p}\ \text{and} \
    \ten{x}_{pj} \cdot \ten{u}_{p} < 0\}
\end{equation}
where, $\ten{x}_{pj} = \ten{x}_p - \ten{x}_j$, $r_{pj} = |\ten{x}_{pj}|$,
$\ten{u}_p$ is the velocity of the boundary particle and $\mathcal{N}_p$ is
the set of neighboring nodes of $p$. In \cref{fig:bm}, the node $2$ satisfies
the conditions in \cref{eq:dir}, and so $J=2$. Using the nearest node index
$J$, and the boundary particle $p$, the direction of motion $\hat{\ten{x}}_r =
-\hat{\ten{x}}_{pJ}$ (where $\hat{\ten{x}} = \ten{x}/|\ten{x}|$).
Thus, the boundary particle position is updated using the following equation
\begin{equation}
  \label{eq:int_solid}
  \ten{x}_{p}^{m+1} = \ten{x}_{p}^m +  (\ten{u}_{p}^{m+1} \cdot \hat{\ten{x}}^m_{r})\hat{\ten{x}}^m_{r} \Delta t,
\end{equation}
where $m$ is the time step.

\subsubsection{Convergence criteria} \label{subsec:conv}

For a perfectly packed distribution of particles, each particle should satisfy
the second condition in \cref{eq:cond1} i.e.\ $\nabla \rho_i = 0$.
However, this would take a lot of computational time. In case of geometries
having irrational volumes like the unit circle (V=$\pi$), one could never
achieve a perfect convergence, given a fixed resolution. Thus, similar to
\cite{jiang2015blue}, the following criteria for convergence is used
\begin{equation}
  \label{eq:conv}
  \frac{\max(u_{i}) \Delta t}{h} < \epsilon
\end{equation}
where, $\epsilon=10^{-4}$ is the tolerance for all our test cases, $u_{i}$ is
the velocity magnitude of $i$\textsuperscript{th} particle, and the maximum
is taken over all the particles.

\subsubsection{Determining the constants and time-step}
\label{subsec:time}

It is important to choose the parameters $\zeta$ and $k_r$ appropriately. It
can be seen that the \cref{eq:backgrounpressure_sph} scales as $O(p_b \Delta
s^{-1})$ and \cref{eq:LJPacc_sph} scales as $O(k_r \Delta s^{-2})$. By
requiring that these forces be of the same order, it can be seen
\begin{equation}
  \frac{k_r}{p_b} = C \Delta s
\end{equation}
where C is an arbitrary constant. To find a suitable value, a 2D lattice of
points is considered. A single particle is perturbed by $\Delta s/4$ in each
direction. A similar procedure is applied on a 3D lattice. A value of $C$ in
the range $0.004 - 0.006$ was found in order to ensure that the two forces are
of the same order.

In order to ensure stability, the damping constant has a form similar to the
one suggested by \citet{colagrossi2012particle} given by $\zeta =
C_{\zeta}/\Delta s$, where $C_{\zeta}$ is set in the range $0.2-0.5$ resulting
in an underdamped system.

The time step, $\Delta t$ is set as in \citet{Adami2013} given by,
\begin{equation}
  \begin{split}
    \Delta t_{p_b} &= 0.1 \frac{h}{p_b} \\
    \Delta t_{\zeta} &= \sqrt{0.1 \frac{h}{\zeta u_{i}}} \\
    \Delta t &= \min(\Delta t_{p_b}, \Delta t_{\zeta})\\
  \end{split}
\end{equation}
where $u_i$ is the velocity magnitude of the $i^{th}$ particle.  We note that
we have used a $p_b=1$ for all our simulations.

\subsubsection{Separating interior and exterior particles}
\label{subsec:inter}

At the end of the simulation, both interior (particles inside the boundary
surface) and exterior particles (particles outside the boundary surface) are
uniformly distributed. The interior particles along with the boundary
particles are extracted and used as solid particles, while the rest of the
particles are used as fluid particles. In order to detect interior and
exterior particles, this simple SPH based procedure is adopted:
\begin{enumerate}[Step 1:]
\item Find the nearest boundary node $j$ to free particle $i$.  Let the normal
  at the point $j$ be $\hat{\ten{n}}_j$.

\item If $\ten{x}_{ij} \cdot \hat{\ten{n}}_j> 0$ then the particle $i$ is
  outside, otherwise it is inside.

\item Step $1-2$ are performed for all particles near the boundary surface.
  For particles that are not near the boundary surface, the neighbors of the
  given particle are checked. If a neighbor is an exterior particle, then it
  too is an exterior particle, otherwise, it is an interior particle. This
  works because all particles near a boundary are already marked.
\end{enumerate}
This process provides a simple method of identification of the interior and
exterior particles. It must be noted this is done after the packing procedure
is completed. This procedure takes much less computational time compared to
the packing procedure.

\subsubsection{Implementation}

The algorithm is implemented using the open-source package PySPH
\cite{pysph2019}. The pseudo-code of the proposed hybrid method is shown in
algorithm \ref{alg:hyb}. The nearest neighbor particle search (NNPS) algorithm
implemented in PySPH \cite{pysph2019} is not a part of the current algorithm.

\begin{sloppypar}
  The algorithm first reads the input using \lstinline{ReadInput} and
  initialize the particles in \lstinline{CreateParticles} as discussed in
  \cref{subsec:geom}. The constants and time step are set in
  \lstinline{SetConstantAndTimeStep}. The first iteration starts with
  \lstinline{UpdateNeighbors} where the neighbor list for every particle is
  created. This is followed by acceleration computation in
  \lstinline{ComputeAccelerations} and integration in
  \lstinline{IntegrateParticles} as discussed in \cref{subsec:pm} and
  \cref{subsec:om}. Nearest free particles are converted and projected to the
  surface in \lstinline{ProjectParticles} as discussed in
  \cref{subsec:pa_add}. This procedure updates the \lstinline{empty_count}
  variable. The iterations continues until convergence condition is true in
  \lstinline{CheckConvergence} as discussed in \cref{subsec:conv}. Finally,
  the particles are separated into internal and external particles in
  \code{SeparateParticles} as discussed in \cref{subsec:inter}.
\end{sloppypar}
\begin{algorithm}[h!]
  \SetAlgoLined
  \KwResult{Coordinates of solids and fluids}
  \code{ReadInput()}\;
  \code{CreateParticles()}\;
  \code{SetConstantAndTimeStep()}\;
  \lstinline{p_freq = 50}\;
  \code{iteration = 0}\;
  \lstinline{empty_count = 0}\;
  \code{converged = False}\;
  \While{\codek{not converged}}{
    \code{UpdateNeighbors()}\;
    \code{ComputeAccelerations()}\;
    \code{IntegrateParticles()}\;
    \If{\code{iteration \% p_freq == 0}}{
      \code{empty_count = ProjectParticles(empty_count)}\;
      \If{\lstinline{empty_count > 4}}{
        \code{converged = CheckConvergence()}\;
      }
    }
    \code{iteration++}\;
   }
   \code{SeparateParticles()}\;
   \caption{Hybrid particle packing algorithm.}
   \label{alg:hyb}
  \end{algorithm}

\subsubsection{Obtaining faster convergence}
\label{subsec:fast}

The algorithm discussed above is the basic form, which adds particles slowly
to the boundary. In case when the boundary surface is smooth and does not
have sharp changes (as discussed in \cref{subsec:geom}), the following
approaches can be taken to speed up the packing process:
\begin{itemize}
\item Using surface point prediction: When the boundary is smooth one may
  project $0.9 N_{s}$ immediately at the start. This perturbs the distribution
  of the particles significantly. If required more particles are projected to
  the boundary while settling the system to an equilibrium.
\item Filter layers near the boundary surface: Conforming particles to the
  boundary surface requires that only some of the particles be moved. Thus,
  one could filter the free particles near the boundary surface and freeze the
  other particles.
\item Reduce projection frequency: One can reduce it slowly once the initial
  projection is complete.
\item Reduce the convergence tolerance: One can potentially reduce the
  tolerance to increase the performance of the algorithm, although this would
  reduce the quality of the distribution of particles.
\end{itemize}
Doing these can potentially reduce the computations by up to a factor of two.
However, we have not performed any of these in the results presented here.

\subsection{Standard Packing}
\label{sec:colla}
In this method, the interior of a 2D object is constructed by using the method
proposed by \citet{marrone-deltasph:cmame:2011}. The object boundary is
represented by a piecewise linear curve (PLC) with normals to the boundary
pointing out of the solid. The first layer of the interior is generated by
moving the PLC points into the body along the normal by $\Delta s/2$ where
$\Delta s$ is the particle spacing. The new PLC is discretized into particles
such that each particle is approximately $\Delta s$ distance apart along the
PLC. The newly added PLC is moved further into the body along the normal by
$\Delta s$ and discretized again. This procedure is repeated until the desired
number of layers of solid particles are generated. It must be noted that this
works only for 2D objects.

Once the dummy layers representing the solid body is created, fluid particles
are placed around the solid particles. The method proposed by
\citet{colagrossi2012particle} is used to pack particles around the fixed
solid particles. In order to initialize the particle position, a grid of
evenly distributed particles is considered and only the particles outside
(defined by the direction of normal) the boundary surface represented by the
PLC are retained. The particles are packed using the number density gradient.
It must be noted that since only the number density gradient is used as a
repulsion force amongst the particles, they are prone to clumping
\cite{swegle1995smoothed, morris1995study}. The particles are subjected to a
damping force to dissipate the energy of the system. Hence the force on any
particle is governed by
\begin{equation}
  \frac{d \ten{u}}{d t} = -\frac{\nabla p_b}{\rho} - \zeta \ten{u},
  \label{eq:cola:force}
\end{equation}
The value of $p_b$ and $\zeta$ is set directly as described in
\cref{subsec:time}. The same SPH discretization is used as done in equation
\cref{eq:backgrounpressure_sph} and \cref{eq:damp_sph}. The convergence
criteria remain the same as discussed in \cref{subsec:conv}.

\begin{algorithm}[h!]
  \SetAlgoLined
  \KwResult{Coordinates of solids and fluids}
  \code{ReadInput()}\;
  \code{CreateParticles()}\;
  \code{SetConstantAndTimeStep()}\;
  \code{converged = False}\;
  \code{iteration = 0}\;
  \While{\codek{not converged}}{
    \code{UpdateNeighbors()}\;
    \code{ComputeAccelerations()}\;
    \code{IntegrateParticles()}\;
    \code{converged = CheckConvergence()}\;
    \code{iteration++}\;
  }
  \caption{Standard particle packing algorithm.}
  \label{alg:cola}
\end{algorithm}

\begin{sloppypar}
In algorithm \ref{alg:cola} the standard packing is described in detail. The
\lstinline{ReadInput} functions reads the points describing the geometry. All
the particles are initialized and the dummy particles are created using method
proposed by \citet{marrone-deltasph:cmame:2011} in
\lstinline{CreateParticles}. The \code{SetConstantAndTimeStep} function sets
the constants and time step as discussed in \cref{subsec:time}. The iteration
starts with creation of neighbor lists for every particle in
\lstinline{UpdateNeighbors}. Then, accelerations are computed in
\lstinline{ComputeAccelerations} using \cref{eq:cola:force} and integrated in
\lstinline{IntegrateParticles} using \cref{eq:velocity}. The iteration
continues until the criteria described in \cref{subsec:conv} is satisfied in
\lstinline{CheckConvergence}.
\end{sloppypar}

\subsection{Coupled packing}
\label{sec:jiang}

\citet{jiang2015blue} proposed a packing algorithm for solid objects both in
2D and 3D in order to sample blue noise. The steps involved are described in
algorithm \ref{alg:jiang0}. A repulsion force which is similar to the one
used in \cite{colagrossi2012particle} is used along with damping. However a
symmetric form of SPH discretization is used given by
\begin{equation}
  \label{eq:backgrounpressure_jiang}
  a_{b,i} = -m_{i} p_{b}  \sum_j m_{j} \left(
  \frac{1}{\rho_{i}^{2}}+\frac{1}{\rho_{j}^{2}} \right) \nabla_i W_{ij},
\end{equation}
\begin{sloppypar}
  \noindent where $W_{ij}$ is the cubic spline kernel function. This is
  computed in \code{ComputeAccelerations} along with an additional force
  discussed later. In the present implementation, a constant background
  pressure, $p_b$ is used. In the original method $p_b= \eta (\rho - \rho_{o})$
  where $\eta$ is a constant. On computing the acceleration, all the particles
  are integrated in \code{IntegrateParticles}.
\end{sloppypar}

Since the particles near the surface lack supporting particles, a large force
acts upon them. In order to keep the particles inside a confined region, the
particles nearer than $0.05 \Delta s$ are converted to boundary particles.
These are integrated similarly as in the case of the hybrid method described
in \cref{subsec:pm}. The boundary particles are projected back to the surface
in every iteration in \lstinline{ProjectParticles} as done in the original
method.

\begin{algorithm}[h!]
  \SetAlgoLined
  \KwIn{\code{particles, max_iter}}
  \KwResult{Packed particles}
  \code{iteration = 0}\;
  \While{\codek{not converged and iteration < max_iter}}{
    \code{UpdateNeighbors(particles)}\;
    \code{ComputeAccelerations()}\;
    \code{IntegrateParticles()}\;
    \code{ProjectParticles()}\;
    \code{iteration++}\;
  }
   \caption{Packing algorithm by \citet{jiang2015blue}.}
   \label{alg:jiang0}
  \end{algorithm}

If one were to only use \cref{eq:backgrounpressure_jiang}, it would result in
more number of particles pushed towards the boundary. In order to counteract
the force on the particles near the boundary, \citet{jiang2015blue} used a
cohesion force proposed by \citet{akinci2013versatile}. The acceleration due
to this force in SPH form
is given by
\begin{equation}
  a_{c, i} = -m_{i} \gamma  \sum_j m_{j} k_{ij} C_{ij} \hat{\ten{n}}_{ij}
  \label{eq:coh}
\end{equation}
where $k_{ij} = 2 \rho_{o}/(\rho_{i} + \rho_{j})$, $\hat{\ten{n}}_{ij} =
\ten{x}_{ij}/r_{ij}$ and $C_{ij}$ is the spline kernel in
\cite{akinci2013versatile} given by
\begin{equation}
  C(q) = \frac{32}{\pi h^{d}} \left \{
  \begin{array}{ll}
    (1-q)^{3}q^{3} & 0.5 < q < 1\\
    2(1-q)^{3}q^{3} - \frac{1}{64}& 0 < q < 0.5\\
    0 & \text{otherwise}\\
  \end{array} \right .
  \label{eq:cs:akinci}
\end{equation}

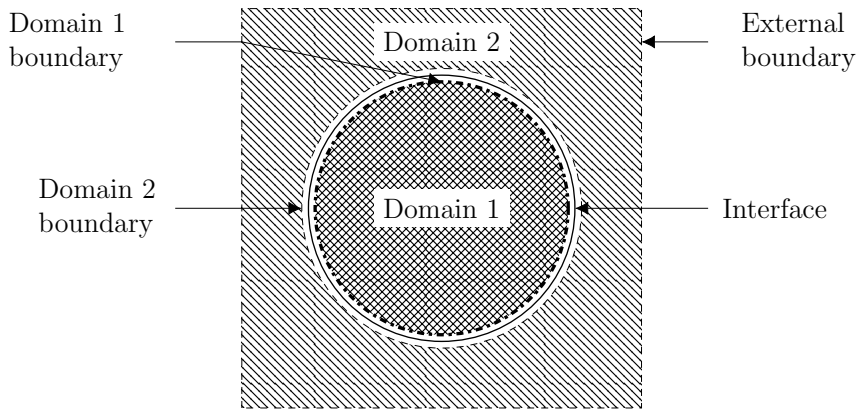
\begin{figure}[ht!]
  \centering
  \resizebox{12cm}{!}{
  \begin{tikzpicture}
    \draw[pattern=north west lines,line width=0.1mm, dashed] (-3, -3) rectangle (3cm,3cm);
    \draw[fill=white, line width=0.1mm, dashed](0,0) ellipse (2.1cm and 2.1cm);
    \draw[line width=0.2mm](0,0) ellipse (2cm and 2cm);
    \draw[pattern=crosshatch, line width=0.5mm, dash dot](0,0) ellipse (1.9cm and 1.9cm);
    \node[fill=white] at (0, 0) {Domain 1};
    \node[fill=white] at (0, 2.5) {Domain 2};
    \draw[<-] (2, 0) -- node[xshift=2cm]{Interface}(4, 0);
    \draw[<-] (-2.1, 0) -- node[xshift=-2cm, text width=2cm]{Domain 2 \\ boundary}(-4, 0);
    \draw[<-] (0, 1.9) -- (-3, 2.5) -- node[xshift=-2cm, text width=2cm]{Domain 1 \\ boundary}(-4, 2.5);
    \draw [<-] (3.0, 2.5) -- node[xshift=2cm, text width=2cm] {External \\ boundary}(4.0, 2.5);
  \end{tikzpicture}
  }
  \caption{Schematic for the coupled packing algorithm. The external region is
    marked as domain 2 and the internal region is marked as domain 1.}
  \label{fig:coupled}
\end{figure}

\begin{algorithm}[h!]
  \SetAlgoLined
  \KwResult{Coordinates of solids and fluids}
  \code{ReadInput()}\;
  \code{DivideDomain()}\;
  \code{CreateParticles()}\;
  \code{SetConstantAndTimeStep()}\;
  \code{Algorithm}\ref{alg:jiang0}\code{(domain1, 5000)}\;
  \code{Algorithm}\ref{alg:jiang0}\code{(domain2, 5000)}\;
  \code{iteration = 0}\;
  \code{converged = False}\;
  \While{\codek{not converged}}{
    \code{UpdateNeighbors(domain1 + domain2)}\;
    \code{ComputeAccelerations()}\;
    \code{IntegrateParticles()}\;
    \code{converged = CheckConvergence()}\;
    \code{iteration++}\;
  }
  \code{SeparateParticles()}\;
  \caption{Coupled particle packing algorithm.}
  \label{alg:jiang}
\end{algorithm}

\citet{jiang2015blue} have not specified a way to choose the value of the
constants in \cref{eq:backgrounpressure_jiang} and \cref{eq:coh}. The values
of $\gamma=20$, $p_{b}=10$ and $h=\Delta s$ are heuristically chosen for all
the simulations.

It must be noted that there is no exterior defined in \cite{jiang2015blue}. In
this work, the exterior is also packed in the same manner as the interior by
moving the boundary surface by $\Delta s/2$ and $-\Delta s/2$ for exterior and
interior respectively. In \cref{fig:coupled}, the interior domain (domain 1)
is enclosed within a thick dashed line and the exterior is within a thin
dashed line (domain 2). Each domain is represented with a different pattern.
Domain 2 having an external boundary has frozen particles outside it. The
solid line represents the boundary surface.

\begin{sloppypar}
The particles are packed in two different passes as described in algorithm
\ref{alg:jiang}. First, the geometry data is read in \code{ReadInput} followed
by particle initialization in \code{CreateParticles}. The free particles are
divided into domain 1 and domain 2 depending upon their location in
\code{DivideDomain}. For particles that are lying in between the two dashed
boundaries of domain 1 and 2, the following is done. If a particle is in the
exterior region (outside the boundary surface) it is moved along the normal by
a distance $\Delta s$ into domain 2. Similarly, particles between the boundary
surface and the domain 1 boundary are moved into domain 1.
\end{sloppypar}

\begin{sloppypar}
In the first pass, domains 1 and 2 are solved separately using the algorithm
\ref{alg:jiang0}. In this case, each domain is unaware of the other. The
boundary particle projection is performed onto the dashed lines of the
respective boundaries. The particles in both the domains are moved for a
predetermined number of iterations at which point the particles reach
equilibrium\footnote{The predetermined iterations is chosen as 5000 currently
  based on the test cases considered.}. This ensures that all particles are
sorted as either interior or exterior and have an interface which they cannot
cross i.e.\ the dashed lines.
\end{sloppypar}

\begin{sloppypar}
In the second pass, when the projection is complete, the particles on the
interior surface i.e.\ the thick dashed line are constrained to move along it.
All other particles are allowed to freely move using
\cref{eq:backgrounpressure_jiang} in \code{ComputeAccelerations} and
\code{IntegrateParticles}. The iteration continues until the convergence
criteria in \code{CheckConvergence} as discussed in \cref{subsec:conv} is
satisfied. The presence of exterior particles eliminates the need for the
cohesion force and is not added once domains 1 and 2 start interacting. Using
this approach, a uniform distribution is obtained both inside and outside the
surface. It is noted that the original implementation was used to sample blue
noise and does not require external particles.
\end{sloppypar}

The proposed algorithms are implemented in the open-source SPH framework,
PySPH \cite{pysph2019}. The present implementation is open source and freely
available with this manuscript at \url{https://gitlab.com/pypr/sph_geom}. All
the results shown in the next section are fully reproducible using a simple
automation framework~\cite{pr:automan:2018}.

\section{Results and discussion}
\label{sec:results}
In this section, all the algorithms discussed above are compared. The
algorithms are first compared for a circular cylinder and a Z shaped wall in
two dimensions. The standard method is limited to two-dimensional domains.
Hence, in three-dimensions, only the coupled and the hybrid method are compared
for an ellipsoid. We also show the particle distribution for a symmetric
airfoil and an arbitrarily shaped geometry using the hybrid method at different
resolutions. In the end, we demonstrate the hybrid algorithm for a
complex-shaped Stanford bunny.

\subsection{Circular cylinder}
\label{sec:cyl}

\begin{figure}[htbp]
  \centering
  \includegraphics[width=\textwidth]{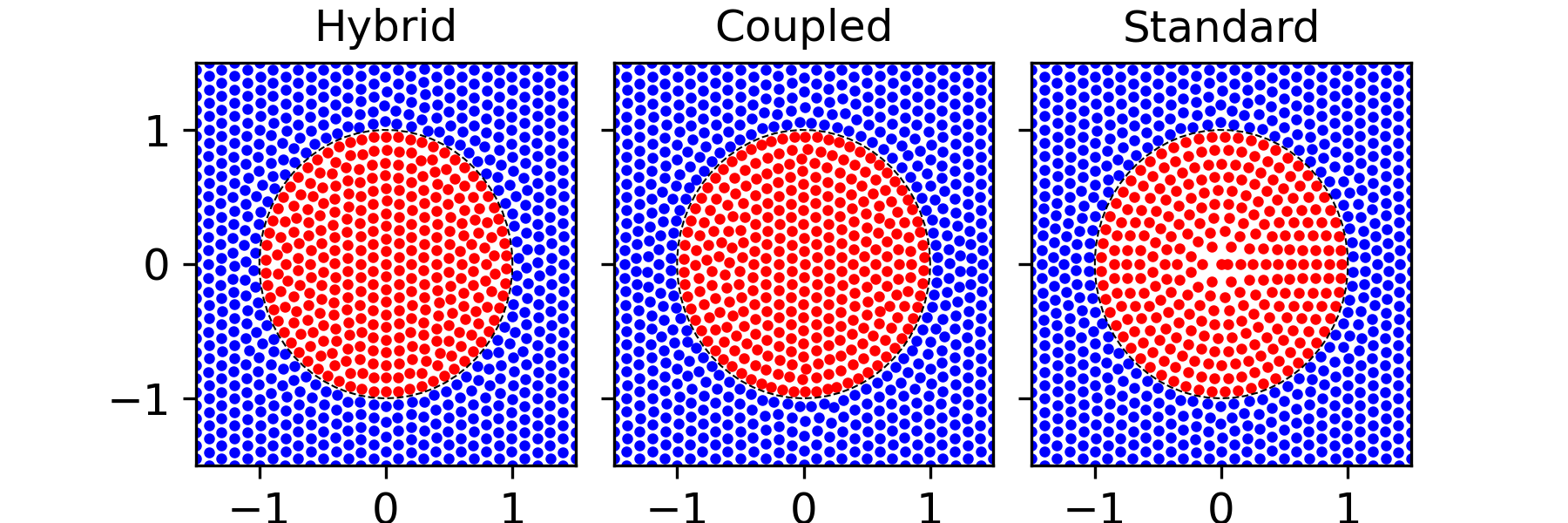}
  \caption{Solid (Red) and fluid (Blue) particles for a circular cylinder for $\Delta s = 0.1$. }
  \label{fig:cyl_sf}
\end{figure}

\begin{figure}[htbp]
  \centering
  \includegraphics[width=\textwidth]{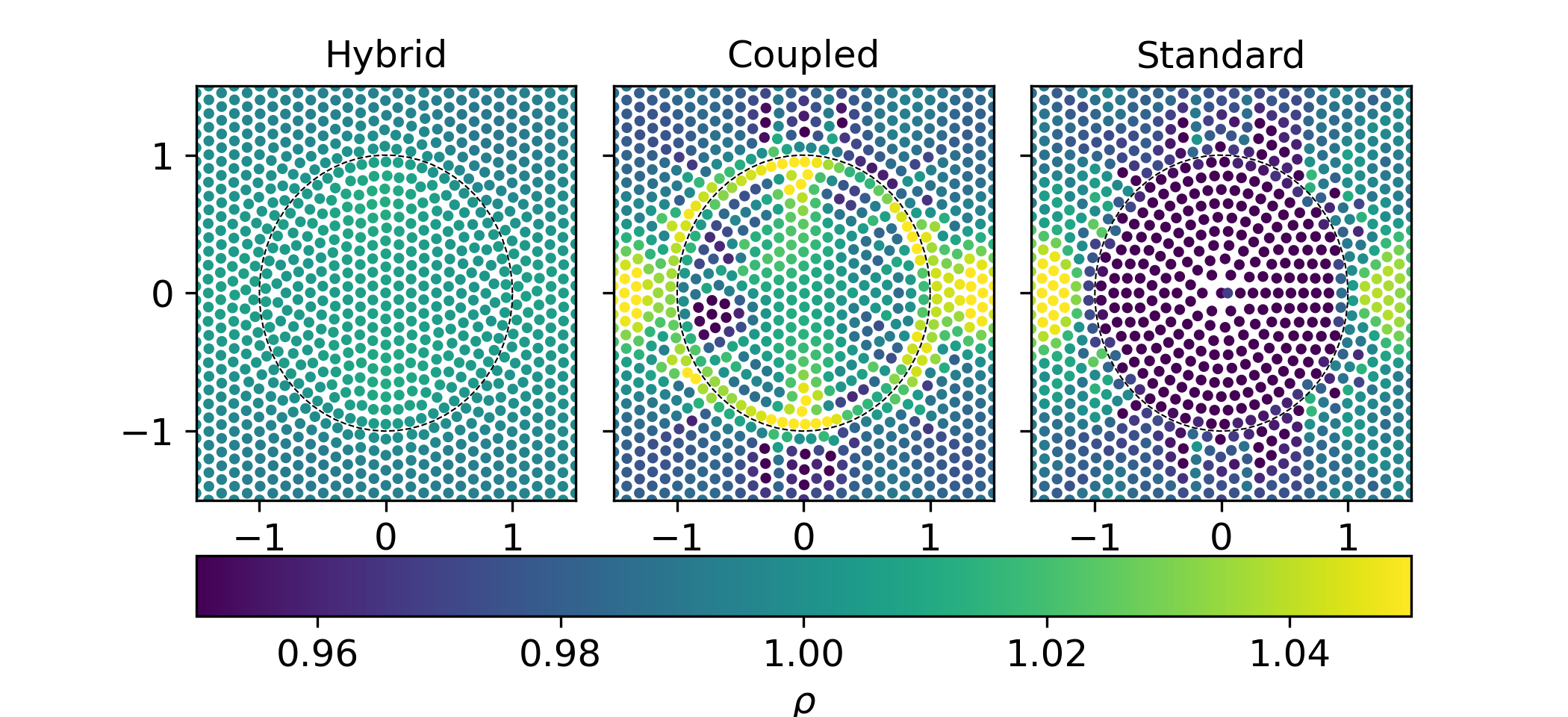}
  \caption{Density distribution of the packed particles for the circular
    cylinder geometry for $\Delta s = 0.1$.}
  \label{fig:cyl_rho}
\end{figure}

\begin{figure}[ht!]
  \begin{subfigure}{0.5\textwidth}
    \centering
    \ifthenelse{\showimages=1}
    {
    \includegraphics[width=\textwidth]{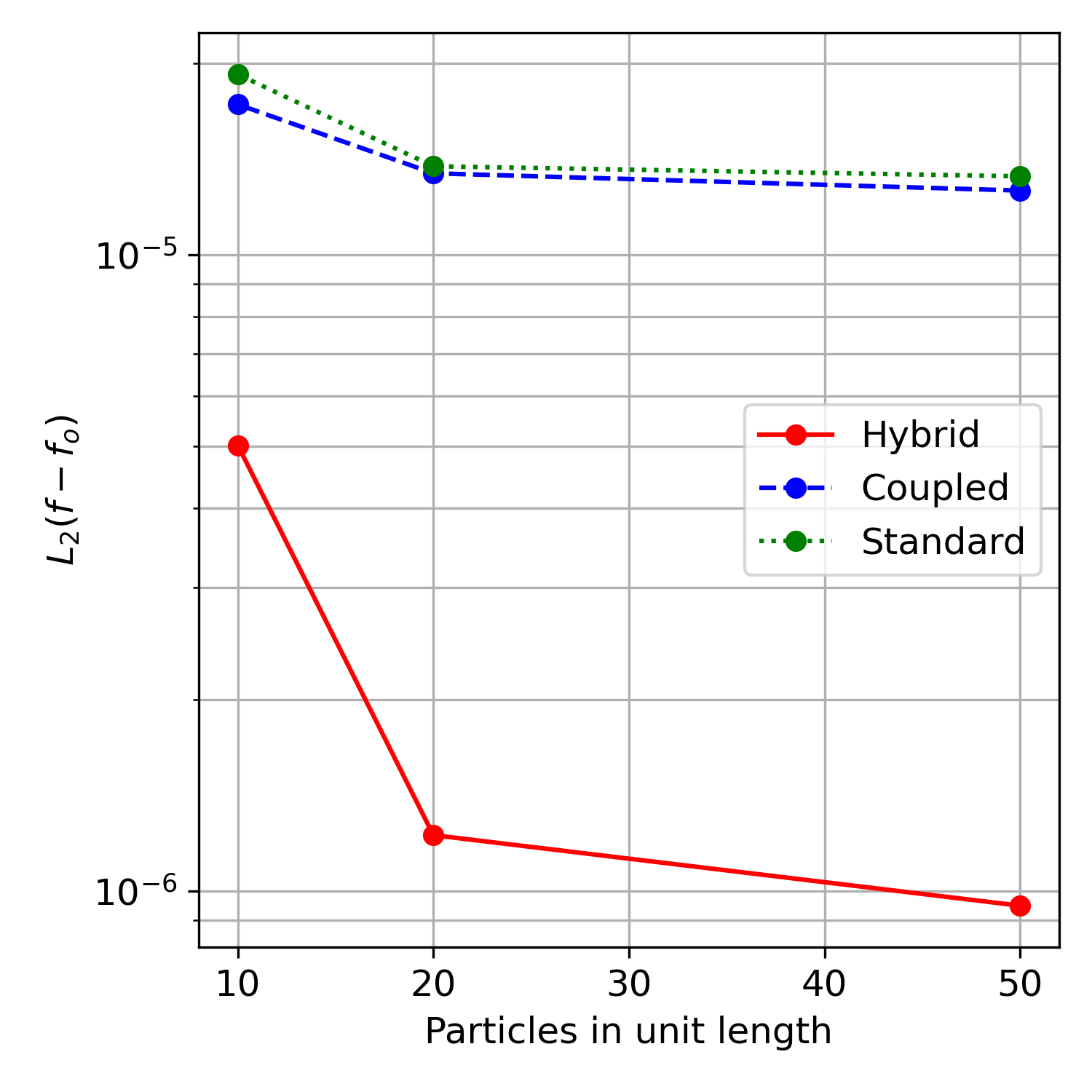}
    }{}
    \caption{Function}
    \label{fig:cyl_f}
  \end{subfigure}
  \begin{subfigure}{0.5\textwidth}
    \centering
    \ifthenelse{\showimages=1}
    {
    \includegraphics[width=\textwidth]{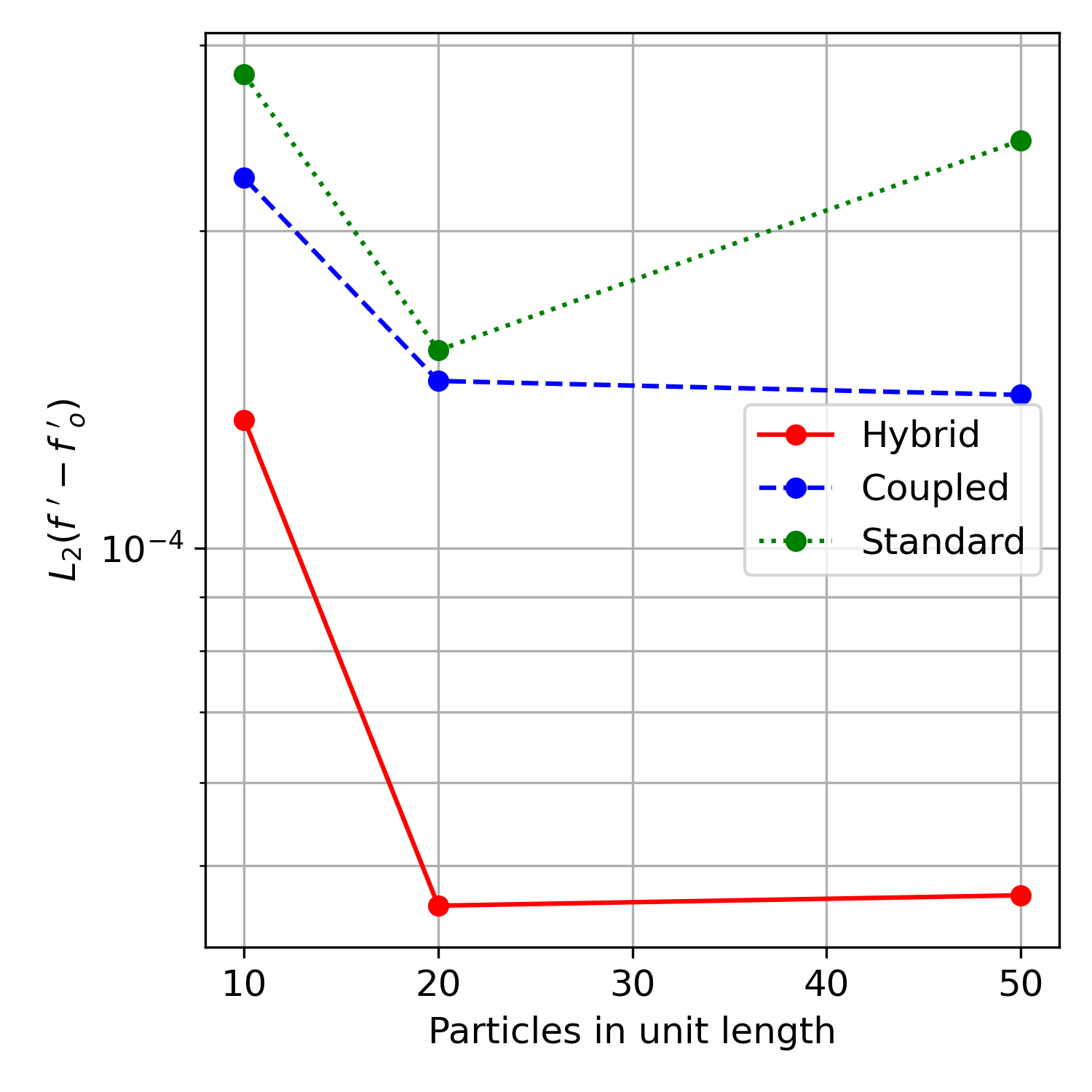}
    }{}
    \caption{Derivative}
    \label{fig:cyl_df}
  \end{subfigure}
  \caption{$L_2$ error for SPH approximation of function and its derivative
    for the circular cylinder geometry.}
  \label{cyl_f_df}
\end{figure}

The flow of an incompressible fluid past a cylinder is a well-known benchmark
problem. In order to obtain a good comparative study, one desires to remove
the effect of the surface irregularities due to the underlying method of
geometry creation. In this test case, the circular cylinder constructed using
all the approaches discussed in \cref{sec:alg} are compared. A cylinder of
diameter $D=2m$ is considered. In \cref{fig:cyl_sf}, the geometry with
particle spacing $\Delta s = 0.1$ made using different methods is shown. It is
clear that the hybrid method produces a uniform particle distribution. In the
case of the coupled method, a large number of particles near the wall surface
is seen. The standard method seems to have uniform particles owing to its
construction. In order to investigate this further, the density distribution
is plotted as shown in \cref{fig:cyl_rho}. The density distribution is
obtained using the well-known summation density. Clearly, the coupled method
shows high density near the surface and the standard method shows a low
density on the solid since $\Delta s$ particle distance is assumed over a
curved surface. The total density variation is 8\%, 16\%, and 2\% for the
standard, coupled and hybrid methods respectively. Thus, it is clear that the
proposed hybrid method shows excellent distribution with a maximum variation
of $2\%$.

The improvement is studied quantitatively by interpolating a $C_{\infty}$
function over the packed particles given by
\begin{equation}
  f(x, y, z) = \sin(x^{2} + y^{2} + z^{2}).
\end{equation}
The function and its derivative is approximated using
\begin{equation}
  <f(\ten{x})> = \sum_j f(\ten{x}_j) W_{ij} \frac{m_j}{\rho_{j}}
  \label{eq:sph_func}
\end{equation}
and
\begin{equation}
  <f_{x}(\ten{x})> = \sum_j f(\ten{x}_j) W_{ij, x} \frac{m_j}{\rho_{j}}
  \label{eq:sph_derv}
\end{equation}
respectively. The $L_2$ error in the approximation is evaluated using
\begin{equation}
  L_2(f-f_{o}) = \frac{\sqrt{\sum_j (f(\ten{x}) - f_{o}(\ten{x}))^{2}}}{N}
  \label{eq:l2}
\end{equation}
where, $f_o$ is the SPH function approximation on a regular mesh of points.
The value of the function is set as per the position of each particle as
$f(\ten{x}_i)$. This is then interpolated onto a regular mesh using
\cref{eq:sph_func,eq:sph_derv}. The same is done for the regular points
themselves to obtain the reference $f_o$ value at each point. The value of $h$
is varied in order to get convergence. A value of $h/\Delta s=1.0$ is taken
for $\Delta s = 0.1$ and linearly varied to $h/\Delta s=1.5$ for $\Delta s
=0.02$. The quintic spline kernel is used for the interpolation. When
comparing the derivatives, only the $x$ derivative is considered. In
\cref{fig:cyl_f} and \cref{fig:cyl_df}, $L_{2}(f - f_{o})$ and
$L_{2}(f^{\prime} - f_{o}^{\prime})$ are shown. In these plots the errors near
the center of the cylinder are not evaluated as they do not affect the flow
and the standard method performs poorly in this region. This change allows for
a fair comparison. Hence, the $L_2$ norm is evaluated only over the points
where $r>0.45 \ D$. It is clear that the hybrid method shows a significant
order of magnitude improvement compared to both coupled and standard methods.
The coupled method is slightly better than the standard method. The standard
method shows large error due to the way in which the surface is represented.

\subsection{Zig-Zag Wall}

\begin{figure}[ht!]
  \centering
  \includegraphics[width=\textwidth]{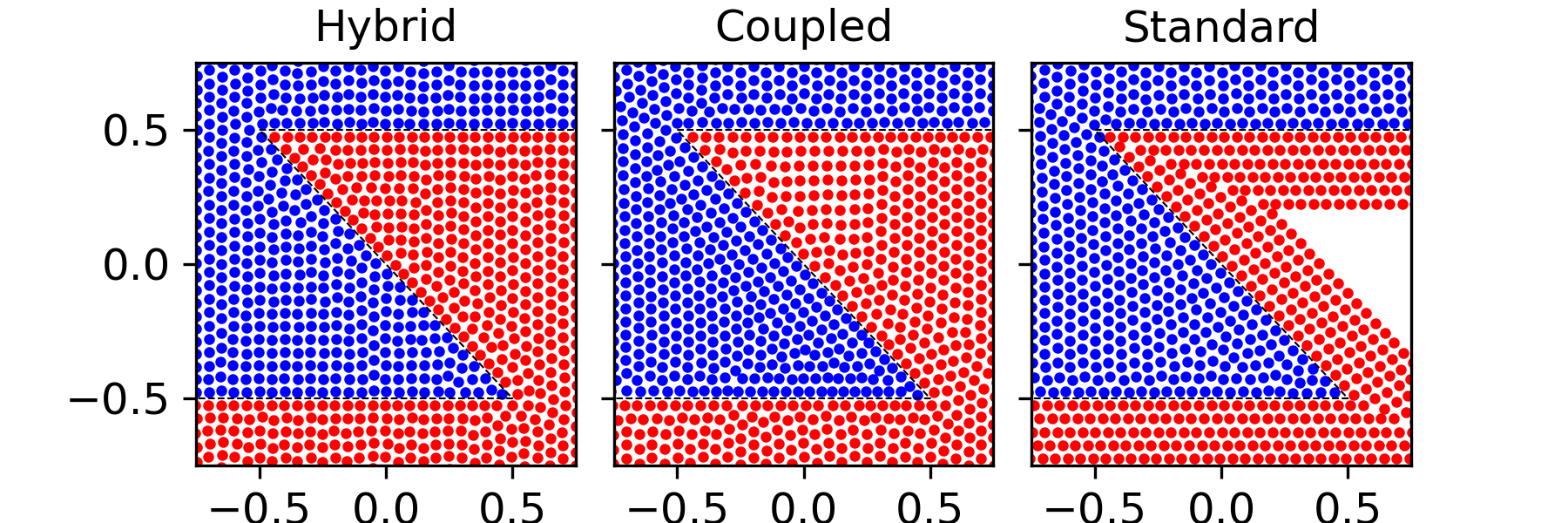}
  \caption{Solids (Red) and fluids (Blue) for the zig-zag wall for $\Delta s = 0.05$.}
  \label{fig:zwall_p_sf}
\end{figure}

\begin{figure}[ht!]
  \centering
  \includegraphics[width=\textwidth]{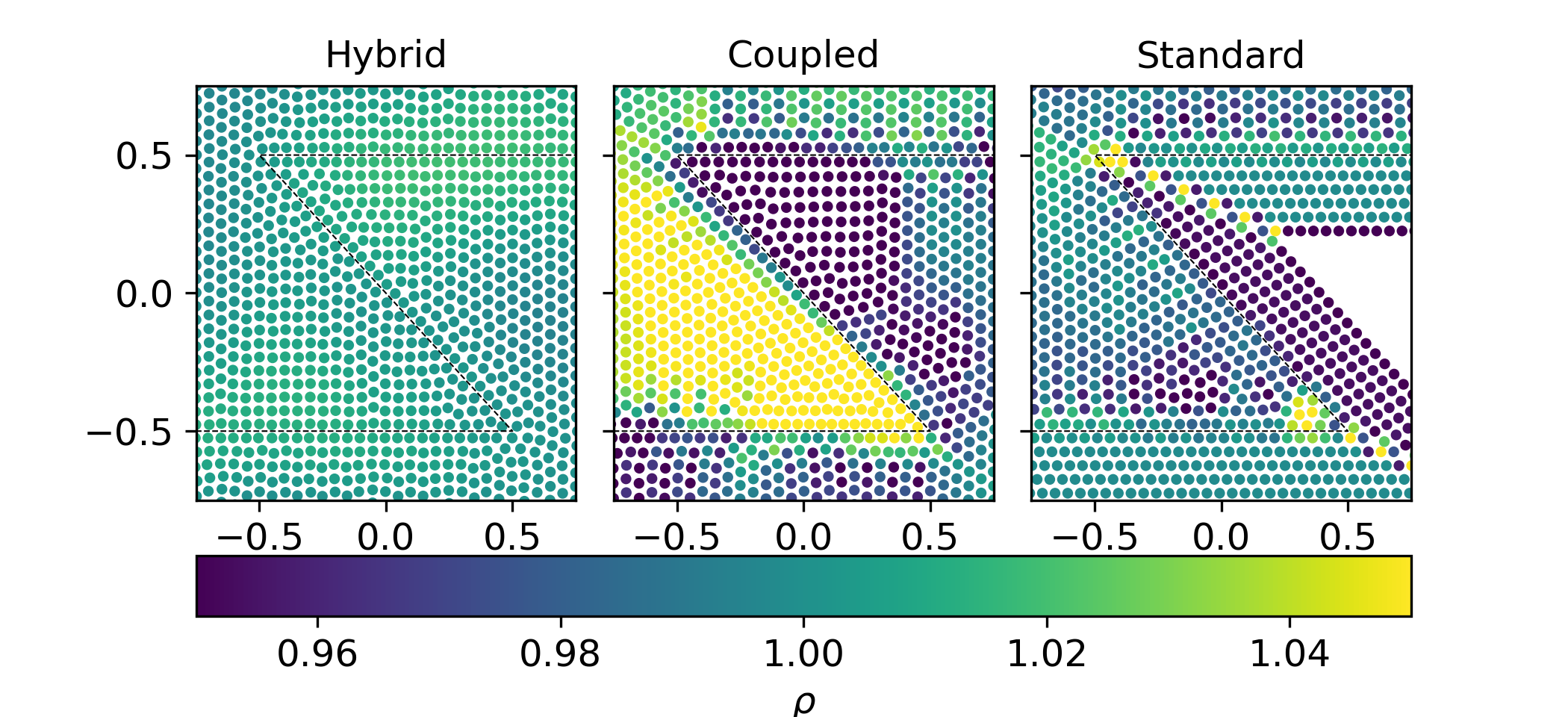}
  \caption{Density distribution for the zig-zag wall for $\Delta s = 0.05$.}
  \label{fig:zwall_p}
\end{figure}

\begin{figure}[ht!]
  \begin{subfigure}{0.5\textwidth}
    \centering
    \ifthenelse{\showimages=1}
    {
    \includegraphics[width=\textwidth]{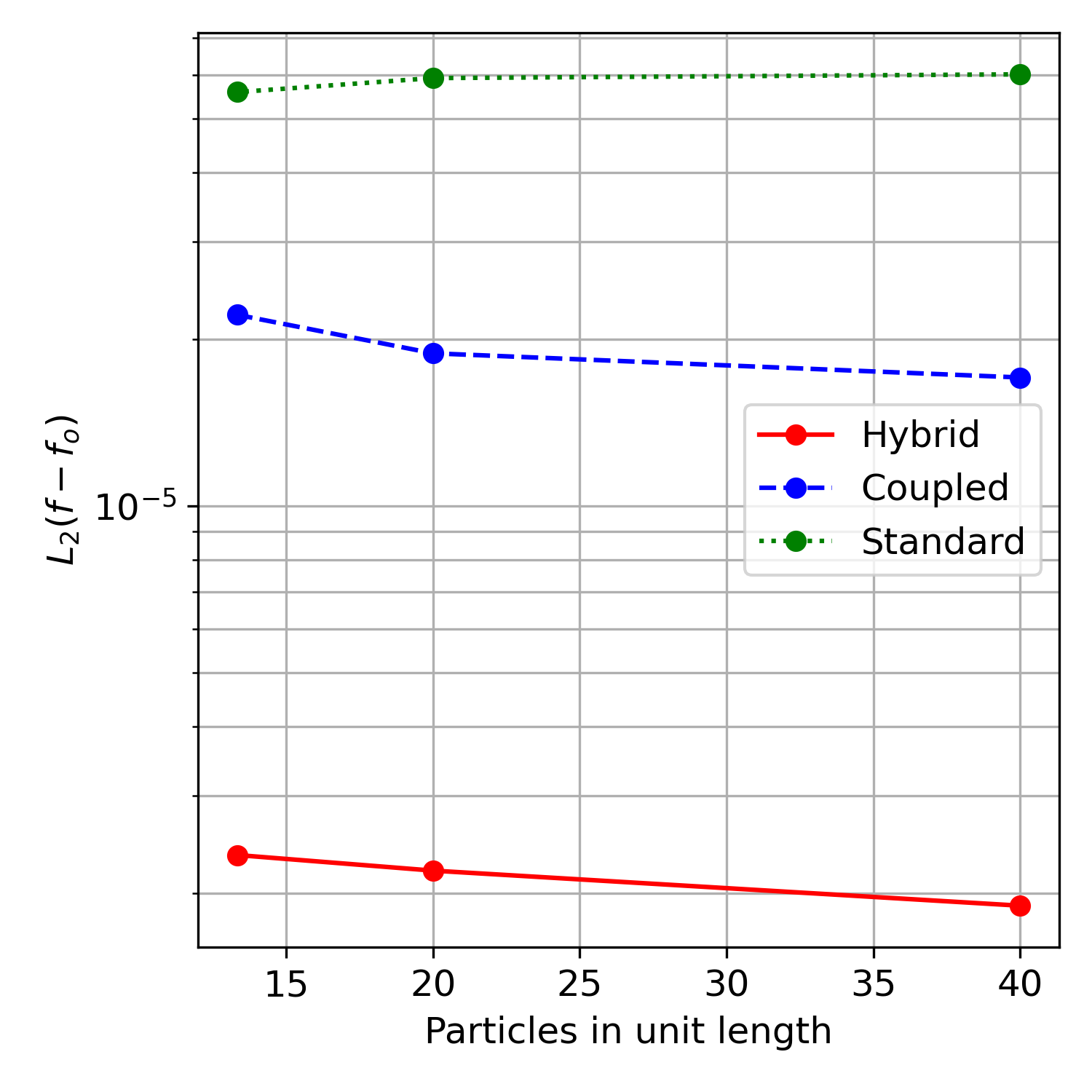}
    }{}
    \caption{Function Approximation}
    \label{fig:zz_f}
  \end{subfigure}
  \begin{subfigure}{0.5\textwidth}
    \centering
    \ifthenelse{\showimages=1}
    {
    \includegraphics[width=\textwidth]{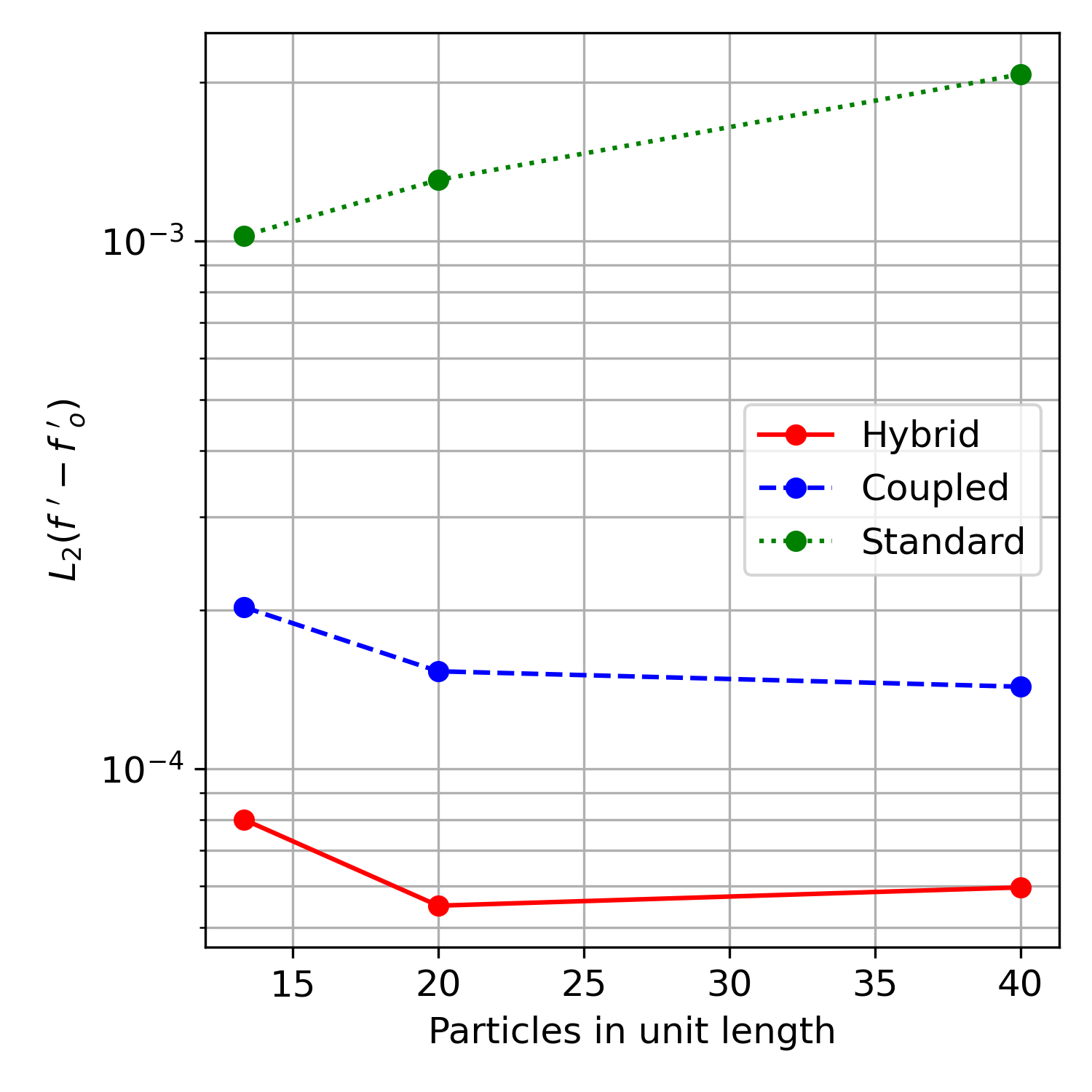}
    }{}
    \caption{Derivative Approximation}
    \label{fig:zz_df}
  \end{subfigure}
  \caption{$L_2$ error for SPH approximation of function and its derivative for the zig-zag wall.}
\end{figure}

The zig-zag wall is one of the test cases proposed by
\citet{marrone-deltasph:cmame:2011} used to demonstrate the $\delta$-SPH
method. They employ the standard packing algorithm in order to generate a
solid body and pack the fluid particles around. In this test case, particles
are packed using all the algorithms and compared. The zig-zag wall is an
excellent test case for packing since it has both concave and convex sharp
edges. In order to generate a solid using the standard method, the corner
nodes are moved along the angle bisector and uniform points are generated
using these points as endpoints. In the other algorithms, these sharp points
are referred as corner nodes and the method discussed in \cref{subsec:pm} is
employed to automatically restrict the motion of these points.

In \cref{fig:zwall_p_sf}, the solid and fluid particles packed with $\Delta
s=0.05$ using different algorithms are shown. It is difficult to conclude
which of these is better. However on looking at the density distribution
computed using the summation density as shown in \cref{fig:zwall_p} one can
clearly see that the proposed method has much less deviation from the desired
density. In the case of the coupled method, higher density is observed at the concave
corner. This occurs since the particles from both sides push towards the
interface at the first pass of the coupled algorithm as discussed in
\cref{sec:jiang}. The standard method shows an uneven variation of density
near the sharp edges which is not desirable. The total density variation is
10\%, 30\% and 3\% for standard, coupled and hybrid methods respectively. It
clearly shows that the hybrid method shows very small density variations
compared to other methods.

A similar analysis over the zig-zag wall is performed as done in the case of the
cylinder. In order to remove the effect of the interior of the solid in the
standard case, the errors are computed only up to a distance of $\Delta s$
inside the wall. In \cref{fig:zz_f} and \cref{fig:zz_df}, $L_2$ norm for the
error in function and its derivative SPH approximation is plotted
respectively. Clearly, the proposed method performs very well as compared with
the other methods.

\subsection{Packing at different resolutions}
\label{sec:resol}

\begin{figure}[ht!]
  \centering
  \ifthenelse{\showimages=1}
  {
    \includegraphics[width=0.45\textwidth]{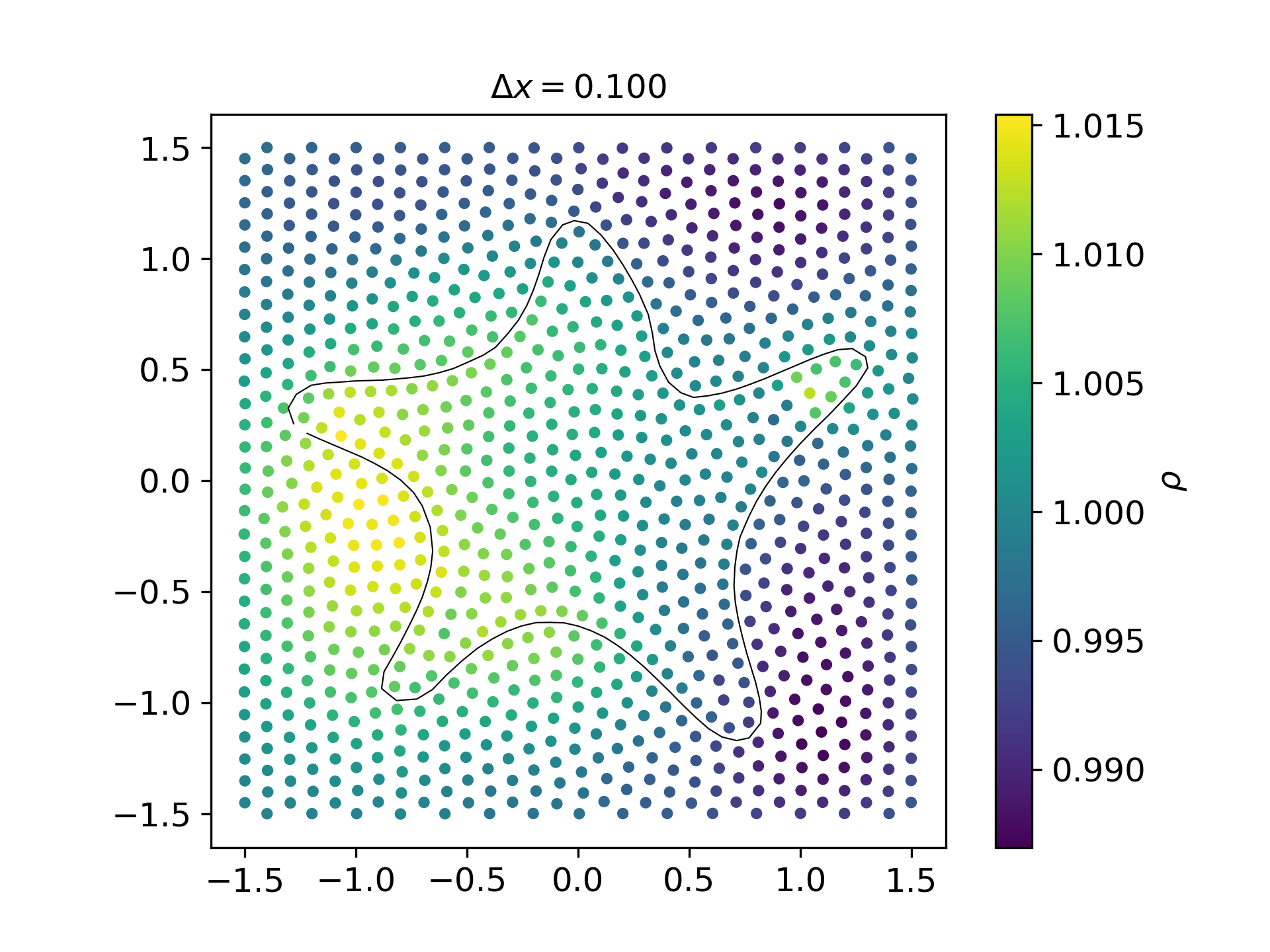}
    \includegraphics[width=0.45\textwidth]{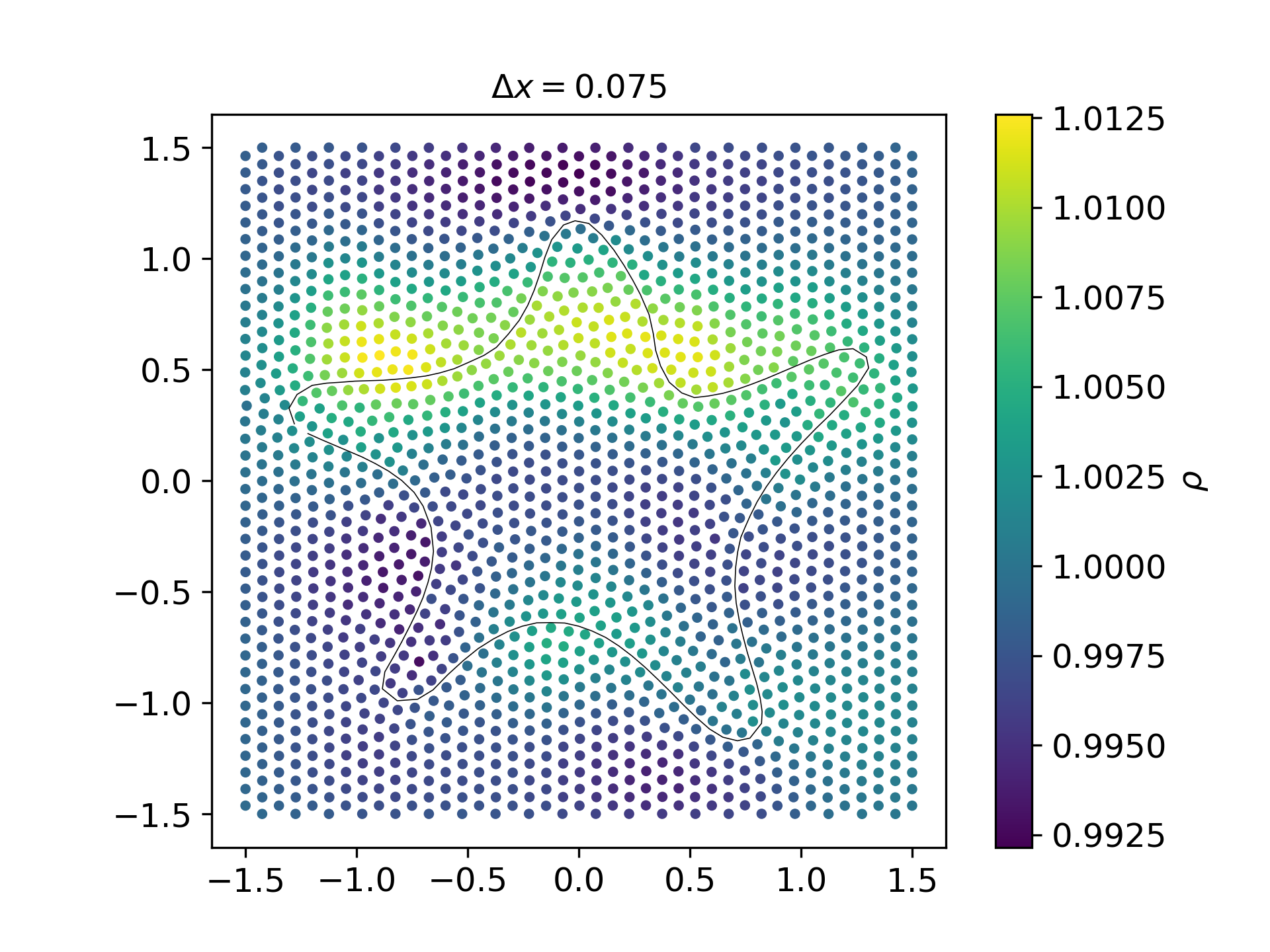}
    \includegraphics[width=0.45\textwidth]{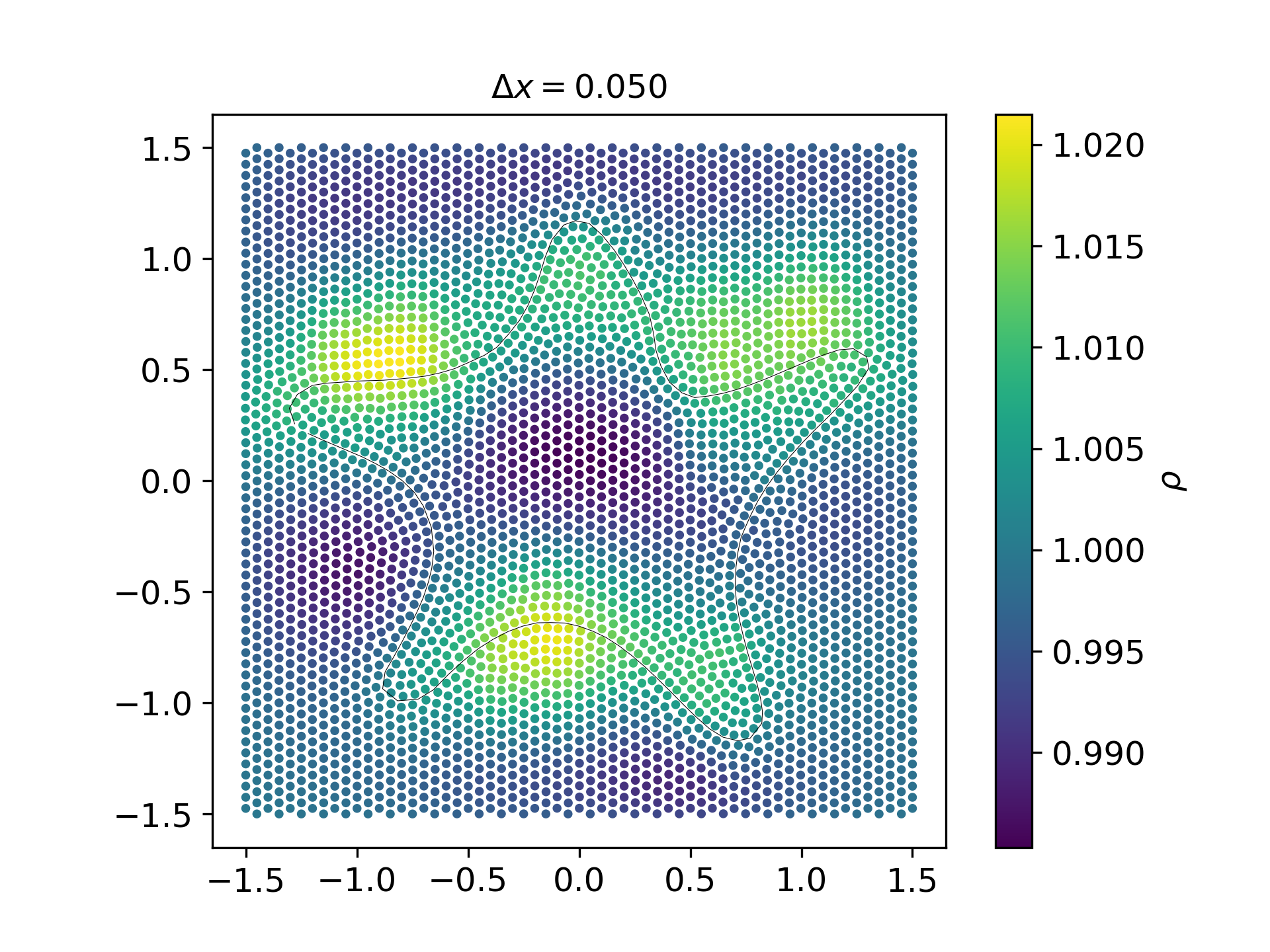}
  }{}
  \caption{Density distribution at different resolutions for an arbitrary shaped object.}
  \label{fig:arb_part}
\end{figure}

In this example, the hybrid algorithm is applied on an arbitrary shaped body. The packing at different particle spacings are shown in
\cref{fig:arb_part}. The particle spacings chosen are $0.05$, $0.075$, and
$0.1$. The particles are placed at $\Delta x /2$ distance away from the
boundary. Clearly, the density distribution is close to the desired value of
$1.0$. The particles conform to the body surface and has a total variation of
density of $2.5\%$, $2\%$ and $3\%$ for particle spacing $0.1$ and
$0.075$ and $0.05$ respectively. This shows that the proposed algorithm is
applicable to complex two-dimensional geometries.

\subsection{Effect of convergence tolerance on the quality}
\label{sec:tol}

\begin{sloppypar}
In this test case, we show that the proposed algorithm can achieve a
high-quality particle distribution for a given tolerance. A symmetric
airfoil, NACA0015 is considered. The end-point at the trailing edge is chosen
as a corner node. Particle packing is performed using the proposed hybrid
method for the spacing $\Delta s= 0.02$.
\end{sloppypar}

\begin{figure}[ht!]
  \centering
  \includegraphics[width=0.7\textwidth]{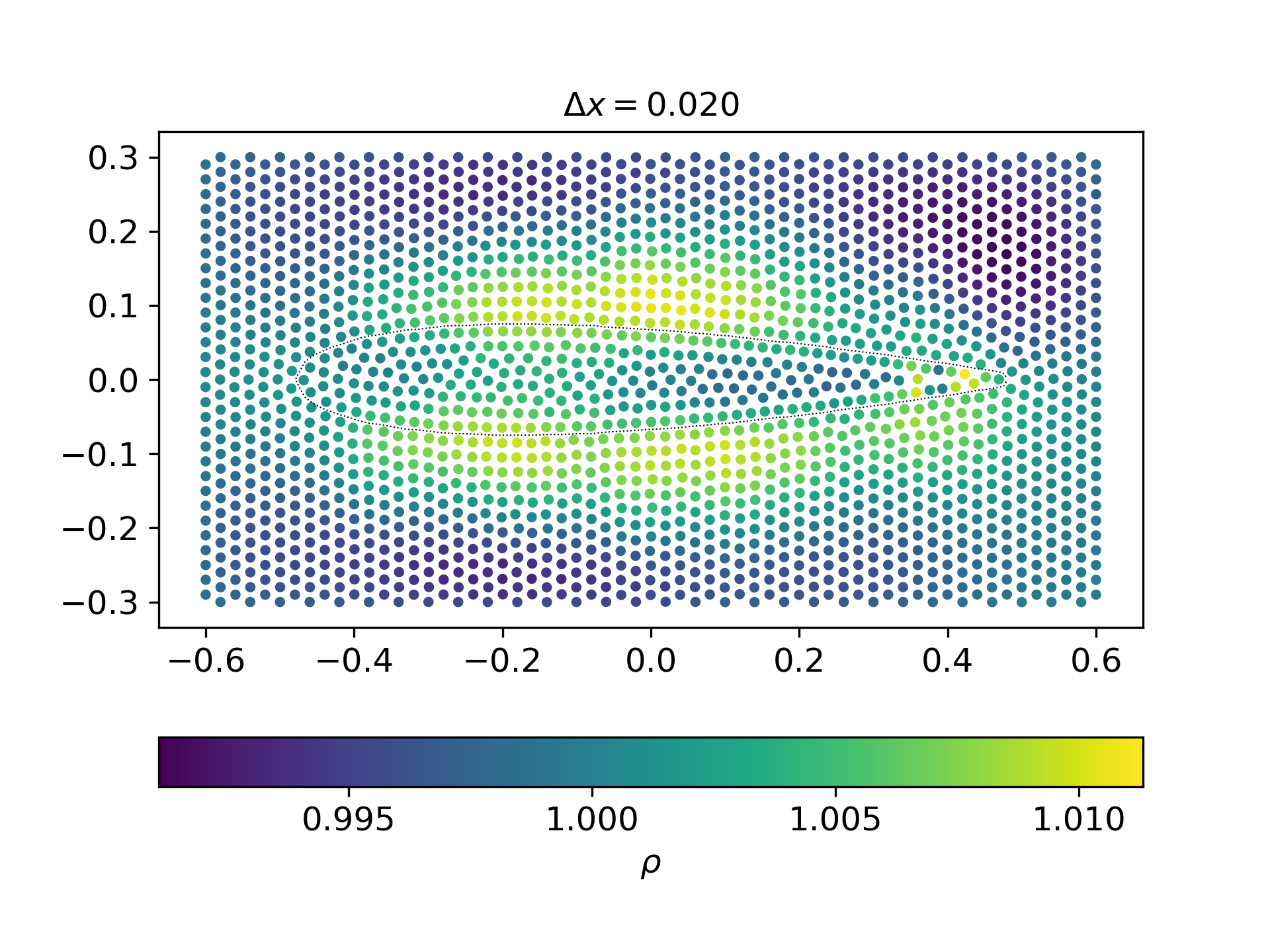}
  \caption{The density distribution for a symmetric airfoil with particle
  spacing, $\Delta s=0.02$ and convergence tolerance, $\epsilon=2.5e^{-5}$.}
  \label{fig:airfoil}
\end{figure}

In \cref{table:tol}, the number of iterations required for convergence and
the $L_{\infty}$ error in the density for different values of the tolerance
required for convergence are tabulated. It is evident from the table that the
$L_{\infty}$ error decreases with the decrease in the tolerance however, the
number of iteration required increases significantly. In the
\cref{fig:airfoil}, we show the high-quality particle distribution achieved
using a lower value of tolerance, $\epsilon=2.5\times 10^{-5}$ in the
convergence criteria. It shows that the proposed method can achieve a
high-quality particle distribution provided a sufficiently low tolerance is
used.

\begin{table}[h!]
  \centering
  \begin{tabular}{cll}
\toprule
Tolerance &  Iterations & $L_{\infty}(\rho_i)$ \\
\midrule
  2.5e-05 &       19641 &                0.011 \\
  5.0e-05 &        6291 &                0.019 \\
  1.0e-04 &        1191 &                0.028 \\
\bottomrule
\end{tabular}

  \caption{The table shows the effect of the change in tolerance for
  convergence on the number of iterations and the $L_{\infty}$ error in the
  density.}
  \label{table:tol}
\end{table}

\subsection{Particle Packing in 3D}

\begin{figure}[ht!]
  \begin{subfigure}{0.5\textwidth}
    \centering
    \ifthenelse{\showimages=1}
    {
    \includegraphics[width=\textwidth]{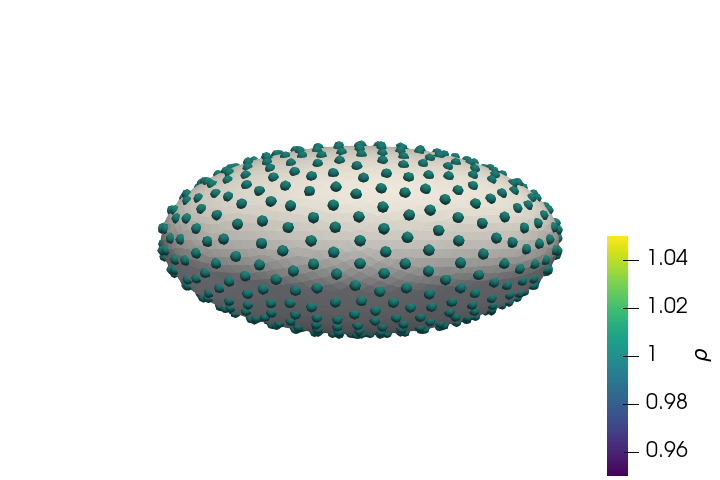}
    }{}
    \caption{Hybrid}
    \label{fig:sph2}
  \end{subfigure}
  \begin{subfigure}{0.5\textwidth}
    \centering
    \ifthenelse{\showimages=1}
    {
    \includegraphics[width=\textwidth]{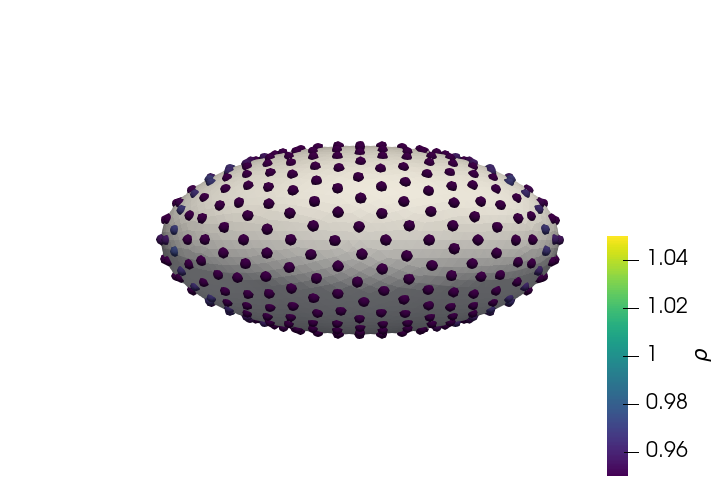}
    }{}
    \caption{Coupled}
    \label{fig:sph1}
  \end{subfigure}
  \caption{Density distribution on the surface of the ellipsoid for $\Delta s = 0.1$.}
  \label{fig:sphere_part}
\end{figure}

\begin{figure}[ht!]
  \begin{subfigure}{0.5\textwidth}
    \centering
    \ifthenelse{\showimages=1}
    {
    \includegraphics[width=\textwidth]{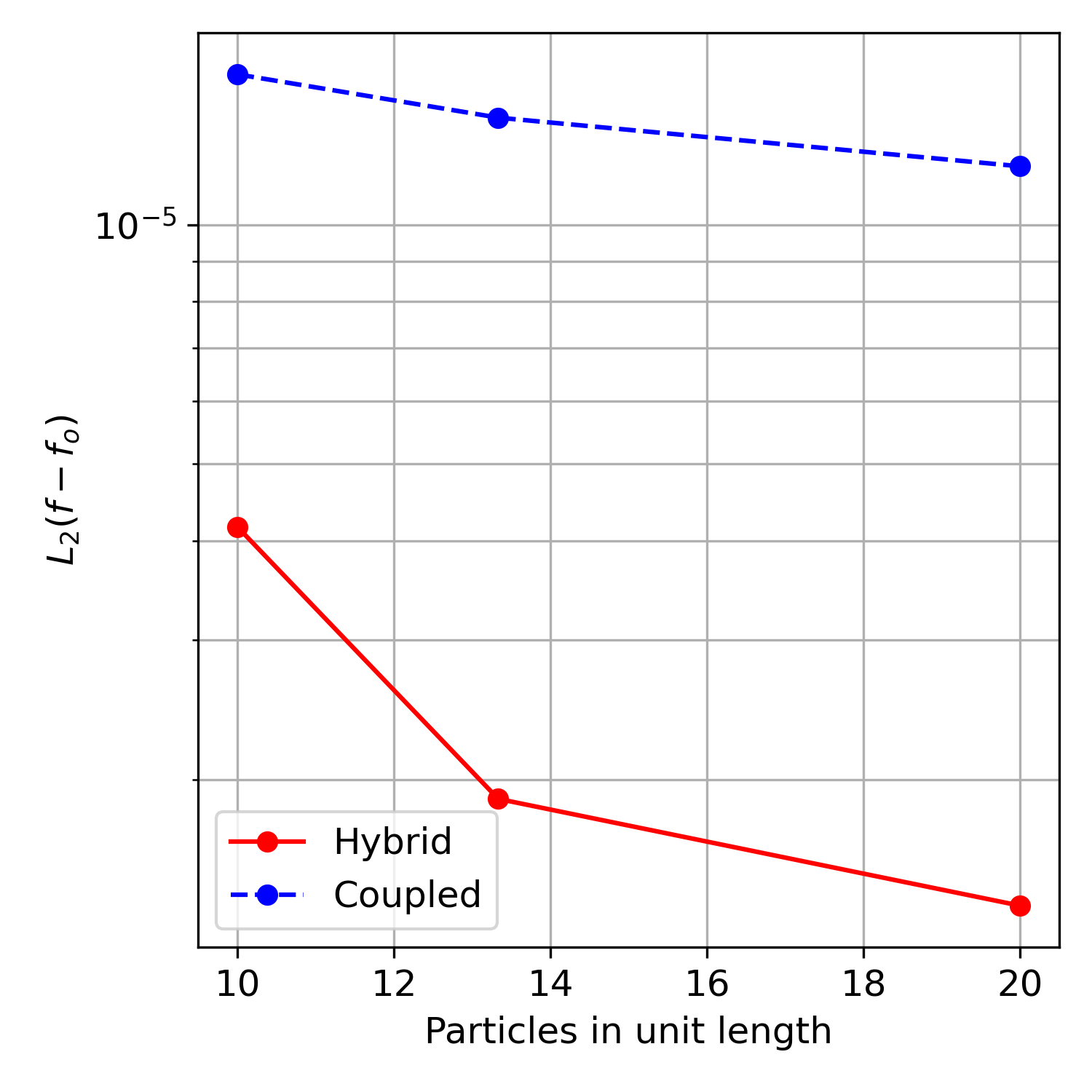}
    }{}
    \caption{Function approximation}
    \label{fig:sphere_f}
  \end{subfigure}
  \begin{subfigure}{0.5\textwidth}
    \centering
    \ifthenelse{\showimages=1}
    {
    \includegraphics[width=\textwidth]{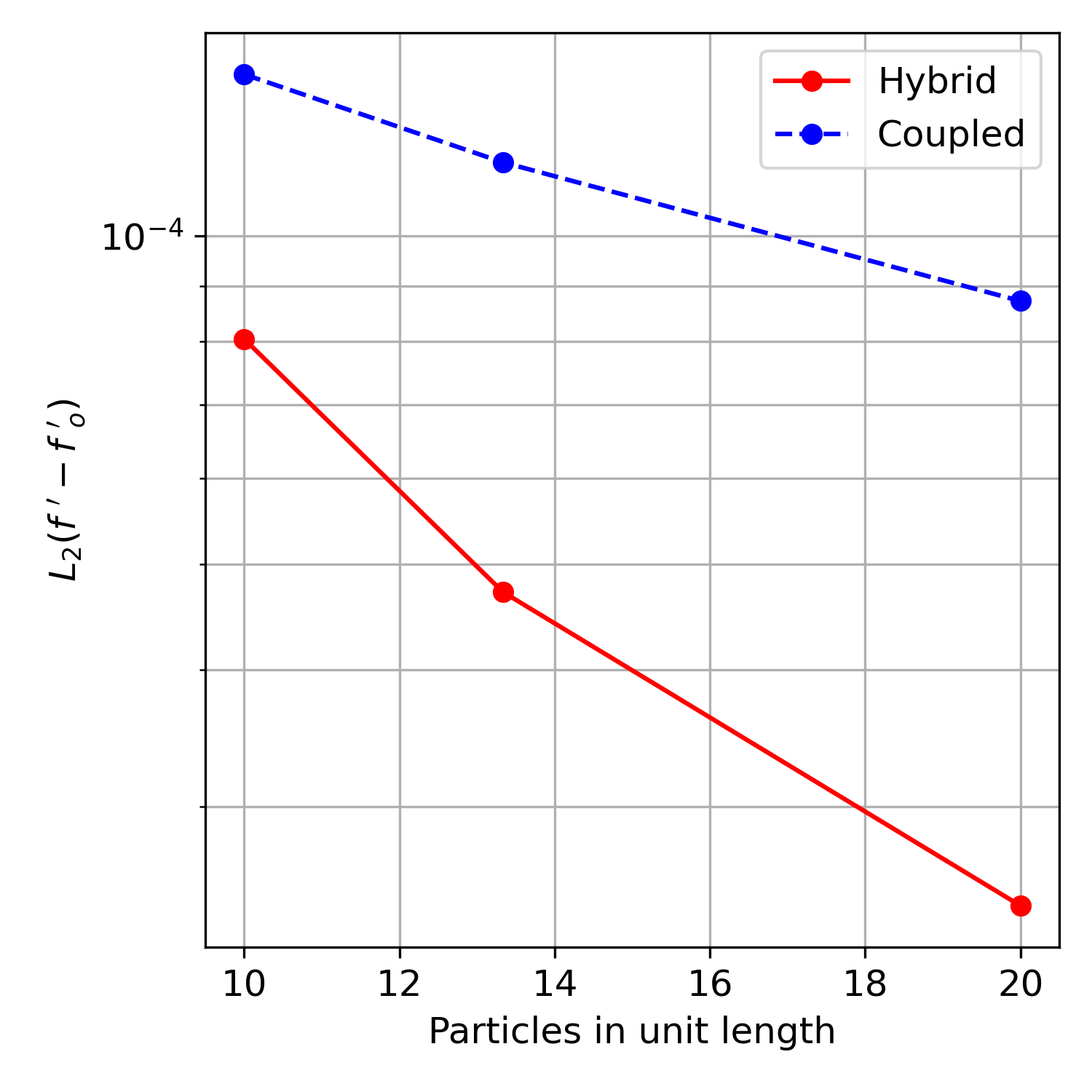}
    }{}
    \caption{Derivative  approximation}
    \label{fig:sphere_df}
  \end{subfigure}
  \caption{$L_2$ error for SPH approximation of function and its derivative for the ellipsoid.}
  \label{fig:sphere_f_df}
\end{figure}
One of the advantages of the proposed algorithm is that it can be easily
extended to a three-dimensional object unlike the standard method. In order to
compare the packing in 3D, particles are packed for a simple ellipsoid. The
ellipsoid has semi-major axis dimensions, $a=1.0$, $b=0.5$ and $z=0.75$ along
$x$, $y$ and $z$ axis respectively. In \cref{fig:sph2} and \cref{fig:sph1},
the packed particles over the surface of the sphere for hybrid and coupled
method are shown respectively. The color of the particles show the density
distribution. In order to show that the particles conform to the surface, the
surface is pulled along the normal by $\Delta s/2$. It is clear that the
hybrid method attains a good distribution of particles resulting in a density
distribution very close to $\rho_{o}=1.0$. The particle distribution using the
coupled method has density near the lower range of the scale. This is due to
the fact that unlike the hybrid method, the coupled method does not project
the required number of particles on the surface. In order to perform a
quantitative analysis, again the comparison of the function and derivative
approximation used earlier is adopted. In \cref{fig:sphere_f} and
\cref{fig:sphere_df}, $L_2$ error in SPH approximation of function and its
derivative is plotted respectively. As can be seen, the proposed method
produces much lower errors at lower resolutions.

\begin{figure}[ht!]
  \centering
  \includegraphics[width=0.45\textwidth,trim={3cm 0 4.5cm 1.5cm},clip]{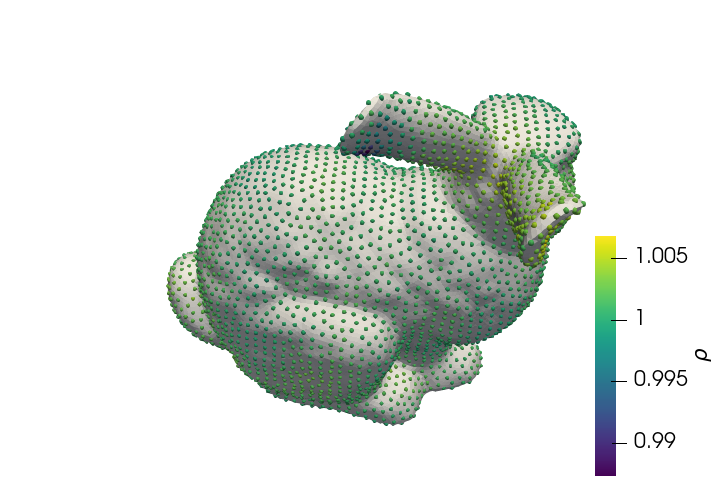}
  \includegraphics[width=0.45\textwidth, trim={3cm 0 4.5cm 1.5cm},clip]{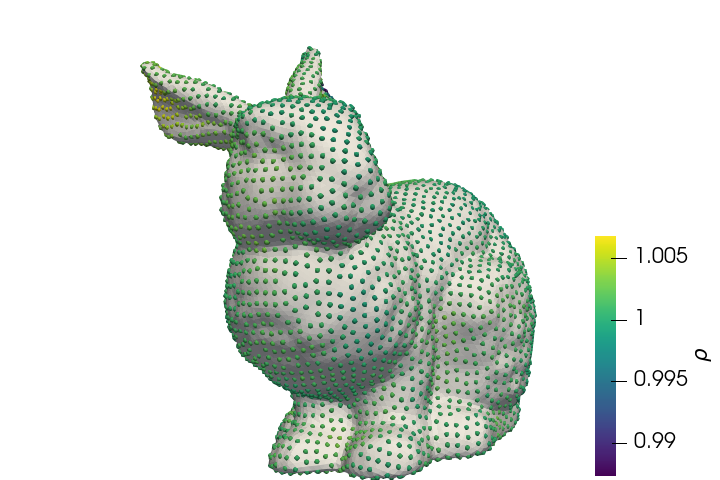}
  \includegraphics[width=0.6\textwidth]{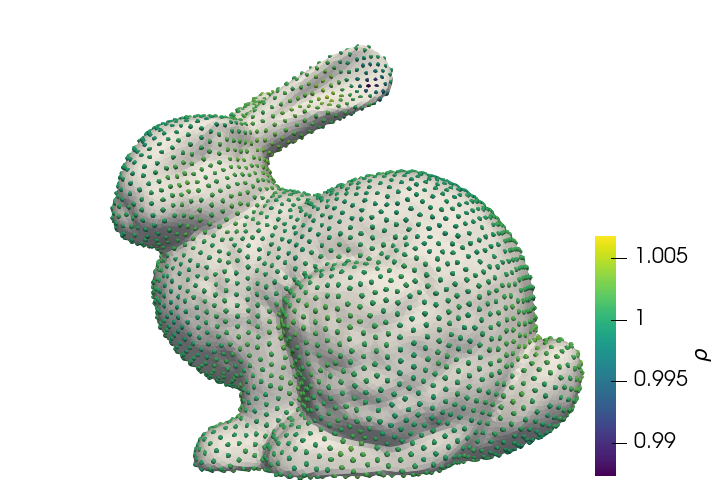}
  \caption{Different views showing particles (colored with density values)
    over the Stanford bunny.}
  \label{fig:bunny}
\end{figure}

In order to show the capability of the algorithm, the proposed algorithm is
also applied to the Stanford bunny geometry, as done by \citet{jiang2015blue}.
The geometry surface required must have outward normals, and the mesh is
corrected using a mesh manipulation tool. A particle spacing of $0.02$ is
chosen. In this case, the intention is to show how well the geometry is
captured, so the surface is not shifted inside. In \cref{fig:sph1}, the
particle distribution over the surface of the bunny is shown. The results show
the applicability of the proposed algorithm to arbitrary shaped 3D objects.
After pre-processing the 3D object can be placed anywhere in the domain, with
the surrounding particles.

\section{Conclusions}
\label{sec:conclusions}

This paper proposes an improved particle packing algorithm for the simulation
of flows involving complex geometries in two and three dimensions. Three
different methods for packing particles around an arbitrary shaped object are
implemented and compared. The standard method which is proposed by
\citet{colagrossi2012particle} along with the solid object construction
proposed by \citet{marrone-deltasph:cmame:2011}. A modified version of that
proposed by \citet{jiang2015blue} which handles both the interior and exterior
of the body. A new method that combines the best features of these methods is
proposed. The proposed method provides an excellent density distribution as a
result of evenly distributed particles. The method is applicable to both 2D
and 3D domains. Unlike the coupled method, no estimation of particles inside
and outside is required. Several benchmark cases are shown which highlight the
accuracy of the proposed algorithm in two and three dimensions. An open-source
implementation of the manuscript is provided and the manuscript is fully
reproducible.

The SPH method is being applied to a wide variety of problem domains involving
complex geometries. While mesh-free particle methods do not require a mesh, one
must still capture the boundaries accurately. However, there are not many
tools to create good quality initial particle distributions. This is necessary
for accurate simulation of incompressible fluid flows with the SPH. The
proposed method is largely automatic and offers researchers the ability to
easily discretize complex geometries using particles in two and three
dimensions. Since the proposed algorithm is based on the SPH, it should be
easy to incorporate into an SPH solver. The code is also open source and hence
researchers could also choose to use our implementation.

In the future, we would like to make it easier to use and integrate with
the main PySPH package. We have only demonstrated our implementation on the
CPU. However, PySPH supports execution on a GPU and we propose to make the
necessary modifications for efficient execution on the GPU. The method
currently works best for a fixed resolution. In the future it would be useful
to explore this in the context of adaptive resolution.

\section*{References}
\bibliographystyle{model6-num-names}
\bibliography{references}

\end{document}